\documentclass[usegraphicx,usenatbib]{mn2e}

\usepackage{epsfig}
\usepackage{amsmath,bm, array,amssymb,aas_macros}
\usepackage[colorlinks]{hyperref}
\newcommand{\Npart}{N_{\mathrm{part}}}
\newcommand{\Nngb}{N_{\mathrm{ngb}}}

\newcommand{\rr}{\left(\bm{r}\right)}
\newcommand{\xx}{\left(\bm{x}\right)}
\newcommand{\ten}[1]{\times 10^{#1}}
\newcommand{\Msun}{\mathrm{M}_{\odot}}
\newcommand{\Mss}{\mathrm{M}_{\mathrm{ss}}}
\newcommand{\leff}{l_{\mathrm{eff}}}
\newcommand{\Nboot}{B}

\title[The Effect of Particle Noise in N-body Simulations of Gravitational Lensing]
{The Effect of Particle Noise in N-body Simulations of Gravitational Lensing}

\author[Rau et al.]
       { S. Rau		$^{1}$ \thanks{Email: rau@mpa-garching.mpg.de},
         S. Vegetti	$^{2}$ \thanks{Pappalardo Fellow},
         S. D. M. White	$^{3}$
\\
$^1$Max-Planck Institute for Astrophysics, Karl-Schwarzschild Str. 1, D-85748, Garching, Germany \\
$^2$ Kavli Institute for Astrophysics and Space Research, Massachusetts Institute of Technology, Cambridge, MA 02139, USA \\
$^3$Max-Planck Institute for Astrophysics, Karl-Schwarzschild Str. 1, D-85748, Garching, Germany \\
}

\begin{document}

\maketitle

\author{Stefan Rau, Simona Vegetti, Simon White}

\begin{abstract}
High resolution dark matter only simulations provide a realistic and fully general means to study the theoretical predictions of cosmological structure formation models for gravitational lensing. 
Due to the finite number of particles, the density field only becomes smooth on scales beyond a few times the 
local mean interparticle separation. This introduces noise on the gravitational lensing properties such as the surface mass density, 
the deflection angles and the magnification. At some small-scale mass limit, the noise due to the discreteness of the N-body simulation becomes comparable to the effects of physical substructures. 
We present analytic expressions to quantify the Poisson noise and study its scaling with the particle number of the simulation and the Lagrangian smoothing size.
We use the Phoenix set of simulations, currently the largest available dark matter simulations of clusters to study the effect of limited numerical resolution 
and the gravitational strong lensing effects of substructure. 
We quantify the smallest resolved substructure, in the sense that the effect of the substructure on any strong lensing property is significant compared to the noise, 
and we find that the result is roughly independent of the strong lensing property. 
A simple scaling relates the smallest resolved substructures in a simulation with the resolution of the N-body simulation.
\end{abstract}


\section{Introduction}
\label{sec:introduction}

The cold dark matter (CDM) paradigm makes two major predictions on the mass structure of dark matter haloes. 
Dark matter haloes at all scales are expected to have a universal mass density profile and to be populated by a large number of mass substructures 
(\cite{Gao_Subhalo_populations_2004,Diemand_cluster_2004,Springel_nature_Millenium_2005, Gao_Phoenix}).
Gravitational lensing provides a unique tool to probe the mass distribution of galaxies and galaxy clusters and therefore to test this model.
At galaxy cluster scale, a number of large gravitational lensing surveys (e.g. CLASH, \cite{CLASH_overview_2012}; SLOAN, \cite{Oguri_weak_and_strong_2012}) 
have accurately determined the shape of the mean mass density profile.
Using a combination of weak and strong lensing data for example \cite{Newman_2009} and \cite{Umetsu_weak_and_strong_2011} 
have measured the mass distribution of galaxy clusters from $\mathrm{kpc}$ to $\mathrm{Mpc}$ scales. 
In general, both in simulations and observations, the central profile of clusters shows a significant amount of 
scatter between different observations (\cite{Sand_2002, Sand_2004,Newman_2011_shallow_cusp}) as well as between different simulations (\citet{Gao_Phoenix}); 
this scatter might be explained in the context of a hierarchical structure formation model where clusters form late and therefore are not fully relaxed objects.

Combining weak and strong lensing is especially important to measure the cluster concentration, defined as the ratio of the virial radius to the radius 
where the density profile has an isothermal slope. 
To date, there seems to be a discrepancy between observations and theoretical expectations, with observed lensing clusters being more centrally concentrated 
than predicted by CDM simulations. This might also be the reason for some unexpectedly large observed Einstein radii (\cite{Zitrin_2011_large_einstein_radii}).
Strong lensing bias might explain these differences (\cite{Comerford_2007_bias_concentration,Oguri_2009_bias_concentraion,Meneghetti_2011_bias_concentration})
due to projection effects for triaxial halos, baryons, or of foreground or background objects.

At present, strong gravitational lensing is the only available tool to detect substructures within the lensing mass distribution beyond the Local Universe and independently of the luminous content. 
Faint and even dark substructures can be detected in lens galaxies either by observing multiply imaged lensed quasars with flux ratio anomalies 
(\cite{Mao_Schneider_flux_ratios_1998_galaxies, Dalal_Kochanek_2002_galaxies, Metcalf_Zhao_2002_galaxies, Kochanek_Dalal_Tests_Substructure_2004_galaxies, Bradac_2002, Fadely_Keeton_Substructure}.) 
or via the gravitational imaging technique on Einstein rings and multiply imaged arcs 
(\cite{Vegetti_2009a, Vegetti_Dark_Substructure_2010, Vegetti_Nature})
While most of the observational and theoretical work that has been done to date is at the scale of galaxies, 
the current search for mass substructure can be extended to galaxy clusters using gravitationally lensed giant arcs. 
The large magnification of these arcs makes them sensitive to substructure masses as small as $\sim 10^8 - 10^9 M_\odot$ that can be detected using the gravitational imaging technique.

Constraining the fraction of mass substructure at the scale of galaxy clusters is also important for understanding the properties of high redshift galaxies. 
Many of the properties (e.g. stellar mass and star formation rate) of high redshift galaxies detected using clusters as cosmic telescopes, 
can be measured accurately only when the source magnification is known. 
However, mass substructure introduces significant fluctuations in the source magnification and needs to be properly accounted for. 
In principle, numerical simulations could be used to quantify the effect of undetected substructure on the source magnification. 
In practice, however, the resolution of the simulation could potentially lead to an underestimate of the substructure fraction in the inner regions, 
while particle noise could mimic the effect of substructure and introduce spurious effects.
In general, from a numerical point of view, when comparing theoretical prediction from numerical simulations and results from observations, 
one should carefully quantify the effect of the limited resolution of the simulation and in particular, the effect of the particle noise.

In this paper, we use the highest available resolution cluster simulations, the Phoenix simulations by \cite{Gao_Phoenix} 
to simulate gravitational lens clusters. Our main goal is to investigate the properties of the particle noise with a focus on multiple lensing properties and gravitationally lensed images. 
The first part of the paper is focused primarily on the quantification of the particle noise for numerical simulations of gravitational lensing, 
while the second part is focused on comparing the effects of particle noise with those of physical substructures.  
In particular, the paper is organised as follows. 
In Section \ref{sec:Numerical_Simulations} we briefly summarize the details of the Phoenix simulations. 
In Section \ref{sec:Lensing_Theory} we will introduce the numerical lensing methods. 
We then quantify the effects of the particle noise on results from N-body simulations in Section \ref{sec:Particle_noise}, considering major lensing quantities 
such as the surface mass density, the deflection angles, the shear, the magnification, the critical lines and the lensed images. 
Section  \ref{sec:scaling} investigates the scaling of the noise with number of particles and smoothing scale. 
In Section \ref{sec:Comparison_of_the_Particle_Noise_with_Substructure}, we compare the effects of noise with those of physical substructures
on each of the lensing properties. This comparison enables us to calculate a resolution limit for the smallest detectable substructures in a N-body simulation of gravitational lensing
as a function of the particle number in the simulation.

\section{Numerical Simulations}
\label{sec:Numerical_Simulations}

In this paper we use the Phoenix simulations by \cite{Gao_Phoenix} which are resimulations of 9 clusters from the Millenium Simulation (\cite{Springel_nature_Millenium_2005}) .
The simulations span masses from $7.5 \ten{14}~\Msun$ for cluster C to $3.3 \ten{15}~\Msun$ for cluster I within virial radii of $1.9$ Mpc and $3$ Mpc respectively.  
All Phoenix clusters are simulated at two resolution levels, Level-2 and Level-4, except for one cluster for which two additional resolution levels, Level-3 and Level-1, are available.
The 3D profiles of the simulated N-body clusters from the Phoenix simulations can be reasonably well fit by two components, a smooth cluster component and additional small-scale substructures.
The smooth component is triaxial and can be fit by either a NFW or an Einasto profile, although an Einasto profile seems to fit slightly better in most cases, 
with an average Einasto shape parameter of $<\alpha> \sim 0.175$.
For a detailed description of the simulations and the parameters of all clusters we refer to \cite{Gao_Phoenix}.

In this paper we focus, as an example, on the cluster E, its properties are listed in Table \ref{tab:phoenix}.
Cluster E is simulated at two different resolutions and has a 
virial mass of $M_{200} \sim 8.1\ten{14} \Msun$ inside $r_{200} \sim 1.9~\mathrm{Mpc}$. 
\citet{Gao_Phoenix} quantify the performance of the NFW and the Einasto fit with a figure of merit function $Q^2$, defined as the 
squared logarithmic deviation $(\ln \rho - \ln \rho^{\mathrm{model}})^2$
averaged over logarithmic radial bins. 
For example for the E halo an Einasto profile with $Q = 0.067$ fits considerably better than a NFW profile with $Q = 0.135$.
Gravitationally bound substructures (subhalos) in the simulation with more than $\sim 20$ particles are identified with the SUBFIND algorithm 
(for a recent comparison of different subhalo finders see \cite{Onions_Subhalo_finder_2012}). 
\citet{Gao_Phoenix} describe the properties of the subhalo population in detail.
In the mass range from $10^{-6} < M_{\mathrm{sub}}/M_{200} < 10^{-4}$ the number of substructures increases with decreasing subhalo mass as $dN/dM \propto M^{-0.98}$. 
The spatial subhalo distribution shows a distinct central core and the subhalos are found mainly at larger distances from the centre of the cluster.

For the lensing simulations throughout this paper we use the same cosmological parameters as in the simulations, 
$\Omega_\mathrm{M} = 0.25$, $\Omega_\mathrm{\Lambda} = 0.75$ and $h = 0.73$.
The N-body simulation stores outputs for redshifts uniformly spaced in $\log a$, where $a$ is the scale factor.
For the analysis presented in this paper, we consider the snapshot at redshift $z = 0.32$ and we place our simulated sources at redshift $z = 2.0$. 
These values are chosen such that the cluster is a relatively efficient lens, and the lensing configuration is comparable to observed strong lensing clusters, 
for example, the median cluster redshift for the CLASH survey is $z = 0.4$ (\cite{CLASH_overview_2012}), while the mean source redshift 
distribution of the SLOAN giant arcs survey peaks at $z = 1.821$ (\cite{SLOAN_source_redshifts}).
\begin{table}
\caption{Properties of the cluster E of the Phoenix simulations.\label{tab:phoenix}}
\begin{tabular}{cccc}
Level 	& $m_{\mathrm{p}}[\Msun]$	& 	$N_{200}$	& 	$\epsilon[\mathrm{kpc}]$	\\ 
4	&	$1.39  \ten{8}$				&	$5.8 \ten{6}$	&	3.84	\\
2	&	$6.06 \ten{6}$				&	$1.3 \ten{8}$	&	0.44	\\
\end{tabular}
\end{table}

\section{Lensing Theory}
\label{sec:Lensing_Theory}

In this section, we describe how we process the simulation data to get a high resolution simulation of the gravitational light deflection
by the N-body cluster.\\ 
During the N-body simulation, snapshots of the positions, velocities and other parameters of the particles were stored 
for multiple redshifts. We use these stored outputs of the N-body simulation for an accurate numerical simulation of the gravitational lensing effect. 
In particular, we use the 3D positions of the particles of the N-body simulation to calculate a smoothed 2D mass density distribution. 
We then use this smoothed particle distribution to calculate the main lensing properties such as the deflection angles $\bm \alpha$, 
the magnification $\mu$, the shear $\gamma$ and the lensed images. 

We adopt the smoothing algorithm commonly used in smooth particle hydrodynamics (SPH) simulations (\cite{SPH_Monaghan}) 
to get a smooth density distribution $\Sigma \xx$ from the $\Npart$ particles of the N-body simulation.
We assign a different smoothing size $l_p$ to each particle.
This particle size is variable and adapted to the local mass density in 3D. 
More specifically, this smoothing size is chosen to be equal to the distance to the $\Nngb$-th neighbour in 3D. 
By doing so, we use the full three dimensional density information to calculate the smoothing lengths of the particles. 
This is computationally more expensive than the equivalent 2D adaptive smoothing, but it results in a 
density map with an enhanced contrast (\cite{2006ApJsmoothing}).
The exact choice for  $\Nngb$ depends on the form of the kernel. Increasing the number of particles in the kernel will 
increase the smoothing and will therefore reduce artificial particle noise, but it will also smooth physical substructures.
In this paper, we use a fourth order polynomial and the distance to the nearest 64 neighbours, for more details see Sec.~\ref{sec:Surface_mass_density}.
In Sec.~\ref{sec:scaling}, we will investigate the dependence of the noise on the number of neighbours.
Once the smoothing length for each particle is calculated, we use two different methods to simulate the gravitational lensing 
by this smoothed N-body mass distribution.
 
The first method is based on a discretization on high resolution grids. 
In order to calculate the surface mass density $\Sigma\xx$, all particles are projected onto a 2D grid
using a smoothing kernel, of smoothing length chosen as described above. 
The surface mass density is then scaled by the critical density $\Sigma_{\mathrm{crit}} = c^2 D_s/(4\pi G D_d D_{ds})$
to get the convergence $\kappa\xx= \Sigma\xx / \Sigma_{\mathrm{crit}}$ where $D_s, D_d$ and $D_{ds}$ are the angular diameter distances from the observer to the source, 
to the lens and from lens to source respectively.
Finally, the gravitational lensing potential $\Psi\xx$ (\cite{Bartelmann}) is calculated by convolving the scaled surface 
mass density $\kappa\xx$ with a logarithmic kernel,
$\Psi\xx = \frac{1}{\pi} \int\!\mathrm{d}^2 x' \kappa\left(\bm{x}'\right) \ln|\bm{x} - \bm{x}'|$,
using a Fast Fourier Transformation (FFT). 
Once the potential has been obtained, the lensing quantities $\bm \alpha, \bm\gamma$ and $\mu$ 
can be calculated by using derivatives of the 1st, the 2nd or combinations of the 2nd order, respectively. 
The derivatives can be done numerically via finite differencing, or in Fourier space directly.
For the highest resolution simulations we project the particles onto two grids of $16384^2$.
Because of zero padding, the size of both grids is increased by a factor of two by the FFT.
The coarse outer grid covers a large area of $(40\cdot 230\arcsec)^2$ around the cluster and includes all external shear components. 
The fine inner grid covers an area of $(230 \arcsec)^2$.  
This corresponds to a maximum resolution of $0.014\arcsec/\mathrm{pix}$ in the inner parts of the halo, which is 
a factor of 5 smaller than the softening length of the original simulations at Level-2. 
Both grids are centred on the cluster.
In order to reduce the computational cost for the bootstrap resamplings described in Sec.~\ref{sec:Particle_noise}, we reduce the grid size to $2048^2$.

The second method for calculating the lensing quantities is the smooth particle lensing method (SPL) introduced by \cite{2007MNRAS.376..113A}.
This method uses Gaussian shaped particles with a smoothing length $l_p$ and 
takes advantage of the linearity of the lensing quantities, $X\rr = \sum_{p = 1}^{\Npart} X_p \rr$,
where $X \in \left\lbrace \bm\alpha, \bm\gamma, \mu,\hdots\right\rbrace$.
As an example the deflection angle at any point on the lens plane is the sum of the contributions of the deflections of all particles
of the N-body simulation. 
This second method is equivalent to convolving the 3D particle distribution with a specific kernel for each of the lensing properties (\cite{2007MNRAS.376..113A});  
for the deflection angles in the $r$ direction, for example, the kernel reads as follows,
\begin{equation}
 W_\alpha(r) = \frac{m_p}{\pi \Sigma_{\mathrm{crit}}} \frac{\exp(\frac{-r^2}{2 l_p^2}) - 1}{r},
\label{eq:alpha_r_direction}
\end{equation}
where $r$ is the distance of each particle $p$ to the evaluated point.
The direct computation of the convolution is very expensive since it requires the evaluation of $\Npart \times N_{\mathrm{ray}}$ values.
However, since the kernel depends on the distance as $1/r$, there exist methods in order to reduce the computational load.
The most popular solution makes use of a tree code to sort and group all particles at large distance and approximate their contribution to any lensing property 
at a given point with a particle of bigger mass (\cite{2007MNRAS.376..113A}). 
In the rest of the paper, we will mostly use the FFT method, but we will also make use of the SPL formalism to quantify the particle noise analytically.

\section{Particle Noise}
\label{sec:Particle_noise}

The mass distribution of any N-body simulation is discretized by individual mass particles. 
This discrete numerical sampling of the underlying mass density distribution introduces shot noise, 
which will affect all the major gravitational lensing quantities.
In particular, the noise level will depend on parameters such as the N-body particle number, the particle size and the smoothing algorithm
that is used to process the N-body particles. 
In this paper we calculate the noise for Poisson distributed particles. 
We are therefore assuming that each particle is sampled from a Poisson distribution with mean and variance one and 
that this sampling is equivalent to the Poisson sampling of an underlying true mass distribution.

The theoretical basis for the expectation value and the covariance analysis for the interpolation of irregular sampled measurements to a smooth map
is covered in a series of three analytic papers by \citet{Lombardi_I_2001, Lombardi_II_2002, Lombardi_III_2003}.
Focusing on gravitational lensing, \citet{Amara_noise_2004} study the effect of particle noise for an analytic singular isothermal ellipsoidal mass model and for a N-body cluster. 
They add artificial 2D Poisson noise to their simulated surface mass density and 
then visually compare the noise in the critical curve and caustic and the magnification of lensed images for several constant smoothing kernel sizes. 
\citet{Bradac_Delaunay_2004} estimate the mean noise on the projected mass density within the critical lines by using 10 bootstrap resamplings (compare Sec.~\ref{subsec:bootstrap})
of the particles of their numerically simulated cluster. They, however, use a smoothing algorithm based on a Delaunay tessellation of the N-body particles for their lensing simulation.
For their analysis of the cusp relation, a relation between the magnification of multiply lensed images close to a cusp (\cite{Schneider_1992_cusp_caustic})
they also compare the particle noise for the Delaunay tessellation smoothing with a more simple Gaussian kernel smoothing algorithm with fixed smoothing size.

\cite{2006ApJsmoothing} suggest adapting the size of the smoothing kernel in order to improve the contrast of the lensing simulation. 
They estimate the noise of their 3D density adaptive smoothing algorithm for a uniform density field, for simulated isothermal ellipsoids and 
for a numerically simulated cluster and compare the results with the results from the simpler 2D adaptive smoothing.
They simulate the particle noise by randomly populating fitted elliptical contours of the smooth halo 
with the same number of particles as the original simulation. 
They conclude that most of the wiggles in the critical line are not due to substructures in the simulation. 
Their simulations of the cusp-caustic relation for the N-body cluster and the Monte-Carlo
re-sampled cluster show that the particle noise produces many high-order singularities.
They also show that a more elliptical projected density is more sensitive to high-order singularities of the caustic. 
This is particularly important for simulating the anomalous flux-ratio problem with N-body simulations.

Although several authors have addressed the problem of the discreteness noise of a N-body simulation and have estimated the effect on some
lensing properties, to our knowledge, to date no one has systematically investigated the implications 
for the simulation of gravitational lensing for very high resolution simulations. 

In this paper we use the smoothing algorithm from \citet{2006ApJsmoothing} and analytically and numerically investigate the magnitude of the effects of particle noise 
on the lensing properties. We extend the investigations of \citet{2006ApJsmoothing} on the particle noise with a focus on different lensing properties and multiply lensed images and compare the noise to
the simulated substructures for a state-of-the-art N-body galaxy cluster simulation.
In this section we first present the two methods that we use to quantify the particle 
noise in the simulation; we then discuss in more detail how the particle noise affects the individual lensing properties 
such as the surface mass density, the magnification, the deflection angles and the lensed images.

\subsection{Bootstrap}
\label{subsec:bootstrap}

In order to simulate the noise numerically, we create $\Nboot$ bootstrapped resamplings of the particles from the N-body simulation.
Each bootstrapped resampling is created by randomly choosing $\Npart$ particles with replacement from the $\Npart$ particles of the simulation.
In this way, some particles will be included more than once, while others will not be included at all.
We then project all particles on a grid and calculate the lensing potential, the deflection angles, the
shear and the inverse of the magnification for each of the $\Nboot$ resamplings. Because this calculation is very time consuming,
we restrict our calculation to $\Nboot = 100$. 
This is sufficient to calculate statistically converged results for the lensing properties that we consider in Sec.~\ref{sec:Surface_mass_density} to \ref{sec:Images}.

To test whether the bootstrapping is a good estimator of the inherent particle noise, we test the bootstrapping method as follows.
We randomly chose $0.01~\Npart$ particles from the high resolution Level-2 simulation with replacement. 
This factor should be large enough to sample the total mass distribution randomly, 
independently of correlated phases, caused by the dynamical evolution during the N-body simulation.
We repeat this 50 times, and then calculate the variance of those 50 randomly sub-sampled clusters. 
To test the validity of the bootstrap method, 
we then randomly choose one additional subset of $\tilde N = 0.01~\Npart$ particles from the original $\Npart$ particles with 
replacement and multiply bootstrap it. 
We, therefore, randomly chose 50 times $\tilde N$ of those last $\tilde N$ particles with replacement. 
The two noise estimates for the subset of $0.01~\Npart$ are identical. 
Therefore we can use the bootstrap method to estimate the particle noise of the original simulation.

For a high resolution lensing simulation, calculating the noise via the bootstrapping technique can be computationally very expensive. 
In particular, most of the computational effort is spent in projecting the simulation particles onto the two-dimensional grids. 
In the following, we present, therefore, a new way to calculate the noise analytically.

We will make use of the SPL method (\cite{2007MNRAS.376..113A}) to derive an analytic expression for the bootstrap noise. 
The SPL method evaluates the contribution of each particle separately, and it is, therefore, ideal to evaluate the noise caused by the discreteness of the N-body simulation.
We assume uncorrelated and Poisson noise for the 3D particle distribution, 
that is, we assume for every particle a Poisson probability density distribution with a mean and variance of one of being included in the resampling. 
For a particle convolved with a kernel, the Poisson noise is smoothed out over the size of the kernel.
The total variance at any point $\bm{x}$ on the lens plane 
is therefore a sum over all the uncorrelated variances of the individual N-body particles,
\begin{equation}
 \sigma^2_{Y}\xx = \sum_{i=1}^{\Npart}W_Y^2\left( |\bm{x} - \bm{x}'_i |,l_i\right),
\label{eq:sigma2_calc_analytical}
\end{equation}
where $Y \in \left\lbrace \kappa,\bm\alpha,\bm \gamma \right\rbrace$ and $ W_Y(|\bm x - \bm x'|,l_p)$ is the appropriate SPL kernel 
listed in the following sections. 
This method calculates the noise directly without the need of numerical bootstrapping. 
However, evaluating Eq.~\eqref{eq:sigma2_calc_analytical} numerically for quantities like for example the deflection angles, is laborious. 
The kernel for the deflection angles is not compact and therefore for any point on the lens plane $\bm x$ one has to sum over the long-range 
contributions for {\it all} N-body particles. In order to effectively evaluate Eq.~\eqref{eq:sigma2_calc_analytical}, one has to use 
some kind of approximation. Two possibilities are, either group distant particles and approximate the effect by a bigger particle in a tree code, 
or to use a two-dimensional approximation (see Sec.~\ref{subsec:Fast_Approximate_Method}).

\subsection{Surface Mass Density}
\label{sec:Surface_mass_density}

The most basic quantity that can be derived from the discrete N-body particle distribution is the smoothed surface mass density. 
In the grid-based approach we derive all other lensing properties via FFT from the projected particle distribution, see Sec.~\ref{sec:Lensing_Theory}. 
In this section, we describe therefore in detail, the smoothing algorithm and the noise properties of the projected mass density distribution.
Simulations of gravitational lensing for a long time used fixed Gaussian kernels to smooth the particles of the N-body simulation
(e.g. \cite{Bartelmann_fixed_SPH_1998, Bradac_2002, Maccio_noise_2006}), 
or adapted the smoothing size if too few particles fell into the kernel (e.g. \cite{Li_semi_fixed_2005}).
A more sophisticated method was then proposed by \cite{Bradac_Delaunay_2004} who used a Delaunay tessellation (\cite{Schapp_Weygaert_Delaunay_2000}) 
to obtain a smooth density distribution from the N-body particles of a numerical simulation.
\citet{2006ApJsmoothing} suggested to adapt the size of the smoothing to the 3D density distribution in order
to improve the contrast of the lensing simulation. In this paper we will use the method from \cite{2006ApJsmoothing}, and use the fully 3 dimensional 
particle distribution to determine the size of the smoothing kernel.
\begin{figure}
\begin{center}
\includegraphics[width=1.0\columnwidth]{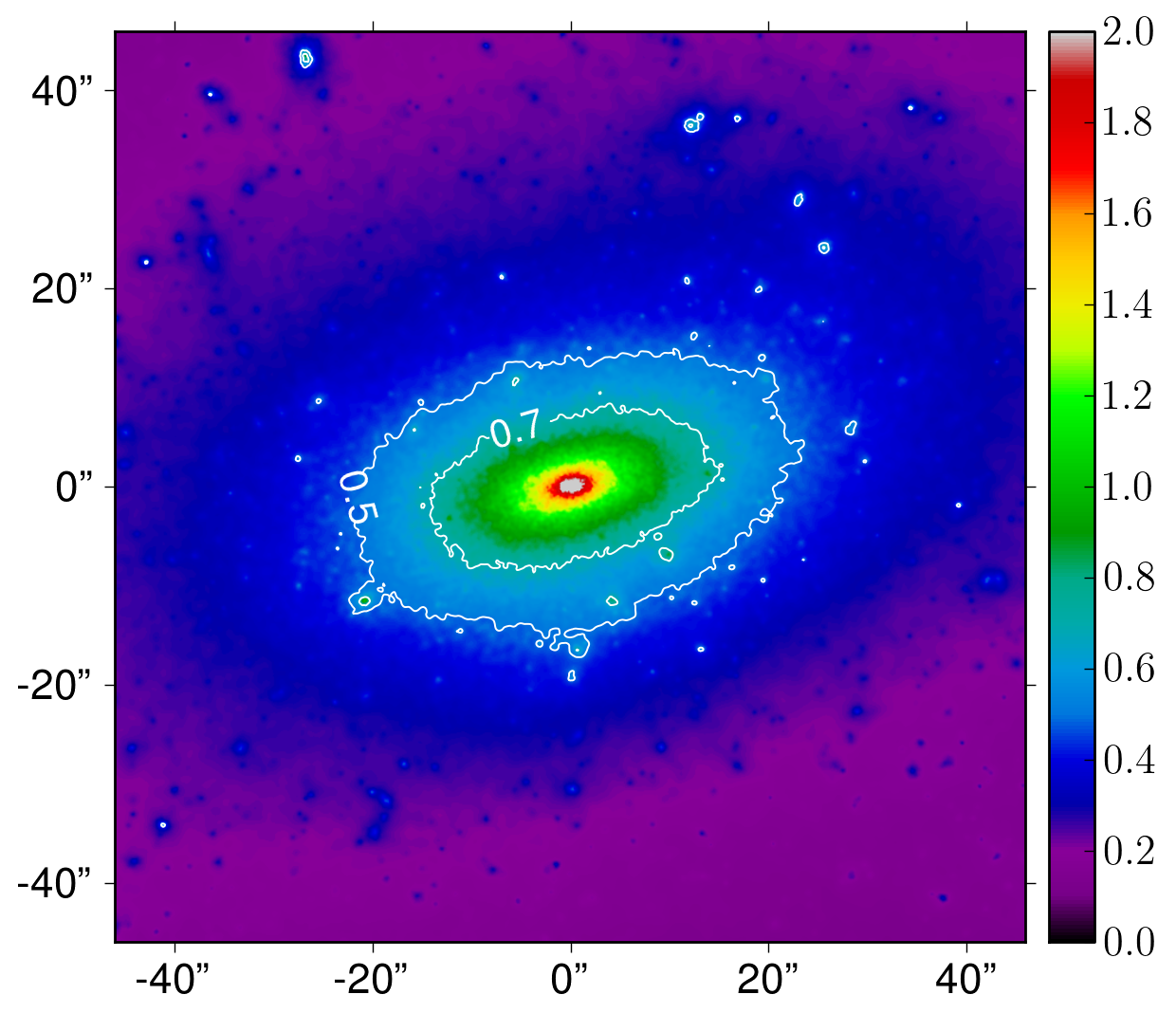}
\includegraphics[width=1.0\columnwidth]{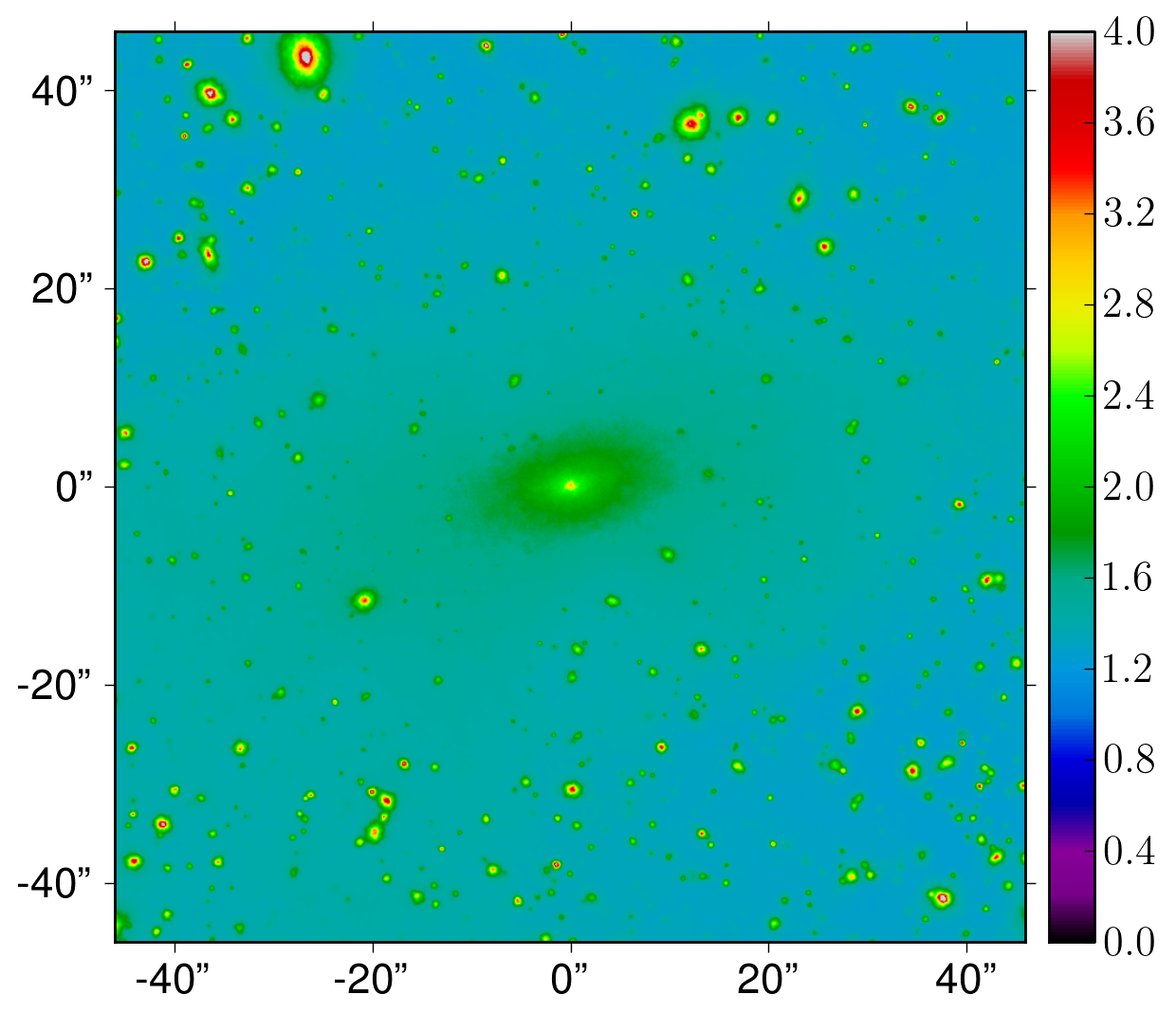}
\end{center}
\caption{
Top panel: Projected convergence of the N-body particle distribution at Level-2 resolution for cluster E. The smoothing is chosen to be equal to the 64th neighbour in 3D. 
Convergence $\kappa$ for the central $(0.55~\mathrm{Mpc})^2$, the critical lines are approximately the same size as the contours at $\kappa = 0.7$. The contours are irregular
because of substructures and the effect of discreteness noise. 
Bottom panel: Particle noise of the convergence $\sigma_{\kappa}\xx$ in percent of the projected scaled surface mass density $\kappa$ 
from 100 bootstrapped resamplings of the particle distribution
\label{fig:mass}
\label{fig:sigma2_sigma_boot_calc}
}
\end{figure}

The 2D surface mass density $\Sigma\xx$ is calculated from the $\Npart$ particles of the N-body simulation. 
Each particle is projected with a normalized kernel,
$\int\!\mathrm{d}^2\bm{x}' W_i(|\bm{x} - \bm{x}_{i}'|, l_i)= 1$ 
in order to obtain a smooth density distribution, 
\begin{equation}
\Sigma\xx = \frac{m_{\mathrm{p}}}{A} \sum_{i=1}^{\Npart}W_i\left(|\bm{x} - \bm{x}_{i}'|,l_i\right).
\label{eq:sigma}
\end{equation}
The size of the kernel, $l_i$, is different for each particle $i$
and is calculated from the distance to the nearest $\Nngb$ neighbours in 3D. 

There are different forms of smoothing kernels, the most widely used 2D functions are Gaussian, 
$W_i(r) = 1/\left(2 \pi \tilde l_i^2\right) \exp\left[-r^2/\left(2 \tilde l_i^2\right)\right]$, 
a simple polynomial, 
$W_i(r) = 3 /\left( \pi l_i^2 \right) \left( 1 - r^2/l_i^2 \right)^2 $  for $r \leq l_i$, 
or the more complicated polynomial (\cite{2006ApJsmoothing,2005MNRAS.364.1105S}),
\begin{equation}
 W_{\Sigma,i}(r) = \frac{80}{14 \pi l_i^2 } 
\left\{\begin{array}{ll}
    1 - 6 (\frac{r}{l_i})^2 + 6(\frac{r}{l_i})^3,      & 0 \le r \le \frac{l_i}{2}, \\
    2 {(1-\frac{r}{l_i})}^3, 				& \frac{l_i}{2} < r \le l_i, \\
    0,                     & \mbox{otherwise.}
\end{array}\right.\;
\label{eq:smoothing_kernel_polynomial}
\end{equation}
For a Gaussian smoothing width $\tilde l_i = \sqrt{103/1120}~l_i$, the effective area 
covered by a particle is equivalent to a particle smoothed by the two other kernels.
The simulation of gravitational lensing does not strongly depend on the form of the kernel.
Since the projection of the N-body particles on the lens plane is the most time consuming step,
we choose the second kernel, which is the fastest to evaluate numerically.

The convergence, the surface mass density in units of the critical density, is shown in the top panel of Fig.~\ref{fig:mass} 
for the Phoenix cluster E at Level-2 resolution.
The figure shows the central part of the halo, the side length is about three times the size of the critical lines. 
The extent of the critical line is comparable to the $\kappa = 0.7$ contours, 
the figure therefore covers the whole region where multiply lensed images occur.
As an example, a second contour at $\kappa \sim 0.5$ is over plotted. 
Both contours are not smooth and show several irregularities. These `wiggles' arise from two different effects. 
The first effect is due to the presence of physical mass substructure in the simulation.
For example, in the central $(92\arcsec)^2$ region, shown in Fig.~\ref{fig:mass}, there are 2597 subhalos identified by SUBFIND with $\Npart > 20$, 
of which 2131 have a $M_{\mathrm{sub}} < 10^{9}\Msun$, 58 have a $M_{\mathrm{sub}} > 10^{10}\Msun$ and
408 have an intermediate mass.
The second effect, which is clearly visible in the top panel of Fig.~\ref{fig:mass}, is related instead to the presence of particle noise.
Even for this very high-resolution simulation both effects are significant and comparable in the inner regions, $r < 20 \arcsec$, of the halo. 
In the following, we will quantify the contribution to the surface mass density fluctuations due to particle noise.

By substituting one of the three smoothing kernels, for example Eq.~\eqref{eq:smoothing_kernel_polynomial}, 
in the expression for the analytic variance, Eq.~\eqref{eq:sigma2_calc_analytical},
we get an analytic expression for the particle noise on the surface mass density,
\begin{equation}
 \sigma^2_{\Sigma}\xx = \frac{m_{\mathrm{p}}^2}{A^2} \sum_{i=1}^{\Npart}W_{\Sigma,i}^2\left( |\bm{x} - \bm{x}'_i |,l_i\right).
\label{eq:sigma2_calc}
\end{equation}

The bottom panel of Fig.~\ref{fig:sigma2_sigma_boot_calc} shows the noise $\sigma_{\kappa}\xx$ on the surface mass density
as a function of $\kappa\xx$. There is a significant difference to a calculation with 2D adaptive smoothing. 
In the 2D case, all line-of-sight N-body particles for a point on the lens plane are assigned the same smoothing length $l$.
For this smoothing we expect the variance of the Poisson noise on the surface mass density to be proportional to the number of line-of-sight particles at each point.
Since the number of particles can be estimated from the convergence $\kappa$ via $A^2\Sigma_{\mathrm{crit}}\kappa\xx / m_p$,
we would expect the bottom panel of Fig.~\ref{fig:sigma2_sigma_boot_calc} to be roughly constant.
In contrast, here, we adapt the smoothing to the local density in 3D, 
and therefore, the noise depends on the integrated particle density distribution along the line of sight and is more complicated than the simple 2D case.
For the Level-2 resolution of the Phoenix simulations in Fig.~\ref{fig:sigma2_sigma_boot_calc} 
the mean noise on the surface mass density is about $1 - 2 \%$ and $<5\%$ even in high density regions.

\subsection{Shear and Inverse of the Magnification}
\label{sec:Shear_and_Magnification}
In the last section we quantified the noise on the surface mass density. 
We will now study the noise on other lensing quantities, namely the shear and the inverse of the magnification, by numerically calculating the lensing potential.
Since the magnification and both shear components are important for weak lensing studies, 
we will also discuss the analytic noise calculation from the particles of a N-body simulation. 
The shear and the scaled surface mass density are related to second derivatives of the lensing potential by  
$\gamma_1 = (\Psi_{,11} - \Psi_{,22})/2$, $\gamma_2 = \Psi_{,12}$  and $\kappa = (\Psi_{,11} + \Psi_{,22})/2$,  where $\Psi_{,12} = \partial_x \partial_y \Psi$; 
from these the magnification can be calculated as $\mu = \left[\left( 1- \kappa \right)^2 - \bm \gamma^2\right]^{-1}$.

\begin{figure}
\begin{center}
\includegraphics[width=1.0\columnwidth]{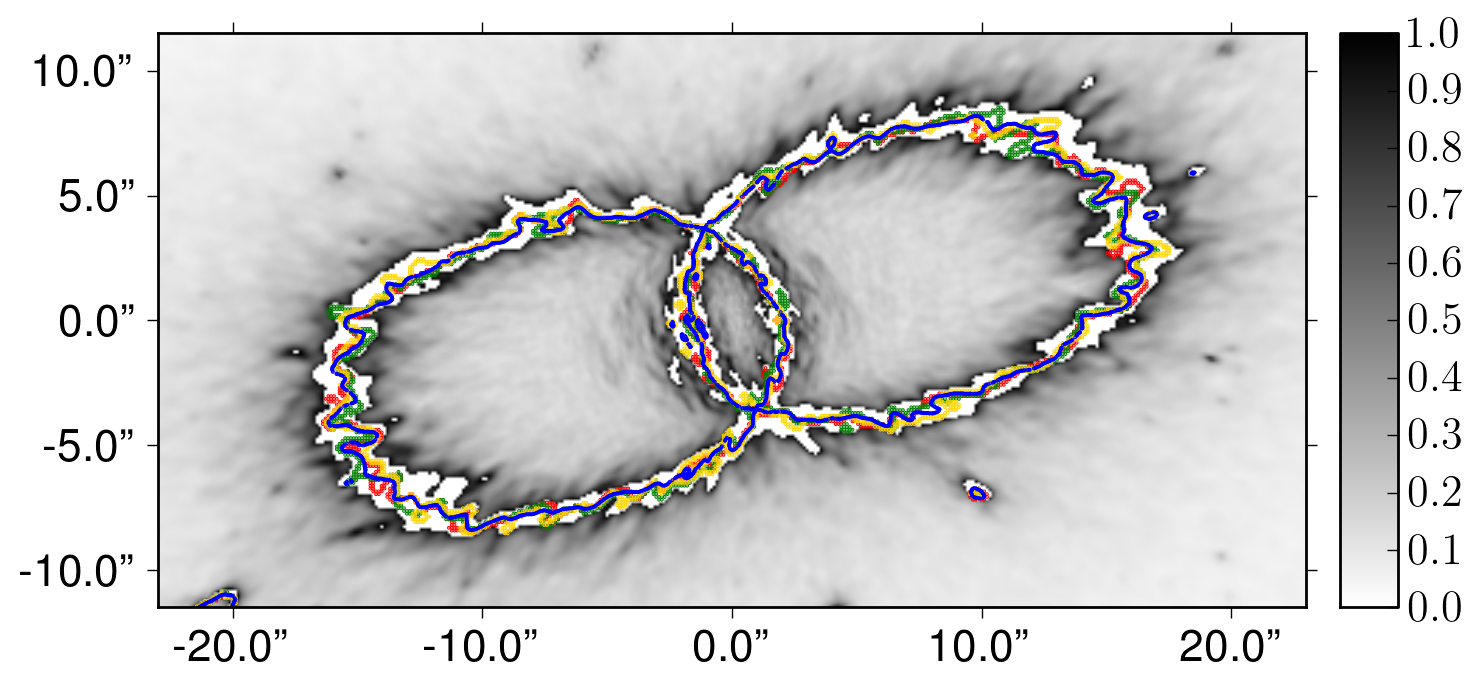}
\end{center}
\caption{Critical lines for different particle noise realisations of the cluster.
Background in grey scale is the particle noise of the inverse of the magnification $3\sigma_{\mu^{-1}}/|\mu^{-1}|$. 
The noise on the inverse of the magnification is cut-off at $3\sigma_{\mu^{-1}} \sim |\mu^{-1}|$ (white band following the critical curves),
this indicates the noise on the critical line. 
}
\label{fig:noise_magnification}
\end{figure}

Using the SPL kernel for the shear from \cite{2007MNRAS.376..113A} we get 
the kernel for the variance of the two components of the of the shear $\bm \gamma = (\gamma_1,\gamma_2)$,
\begin{equation}
 \bm W_{\sigma^2_{\bm\gamma}}\left(r,l_p \right) = \frac{m_p^2 G^2 }{\Sigma^2_c\pi^2 r^8} 
  \left\lbrace \frac{r^2}{2 l_p^2} + \left[ 1 - G\left(r_p,l_p\right) \right] \right\rbrace^2
\begin{pmatrix}
\left(x^2 - y^2 \right)^2
\\
4x^2y^2 
\end{pmatrix}
\label{eq:shear_kernel}
\end{equation}
where $G(r_p,l_p) = \mathrm{exp}\left[-r^2/\left(2 l_p^2\right)\right]$ and $r^2 = x^2 + y^2$. 
We use this kernel to calculate the variance of the two shear components,
$ \bm\sigma^2_{\bm\gamma} = \sum_{i=1}^{\Npart} \bm W_{\sigma^2_{\bm\gamma}}\left(\bm x_i - \bm x,l_i \right)$.
Due to the term $G^2$, the support of the kernel \eqref{eq:shear_kernel} is more compact than the Gaussian smoothing kernel. 
Therefore Eq.~\eqref{eq:shear_kernel} can be easily evaluated by direct numerical summation. This provides a fast and easy way to calculate the
particle noise on the two shear components. 

Taylor expanding the inverse of the magnification allows us to 
derive the noise on the inverse of the magnification due to noise on $\bm\gamma$ and $\kappa$, 
\begin{equation}
 \sigma^2_{\mu^{-1}} = 4 (1 - \kappa)^2\sigma^2_\kappa + 4 \gamma_1^2 \sigma^2_{\gamma_1} + 4 \gamma_2^2 \sigma^2_{\gamma_2}
\end{equation}
As an example, Figure \ref{fig:noise_magnification} shows the $3\sigma$ particle noise on the inverse of the magnification $3\sigma_{\mu^{-1}}/|\mu^{-1}|$ as a grey scale background. 
In order to enhance the effect, the noise level is set to white for all pixels where $3\sigma_{\mu^{-1}} > |\mu^{-1}|$ close to the critical line.
This indicates the width of the `wiggles' of the critical line due to discreteness noise in the region close to the critical line where $\mu^{-1} \rightarrow 0$. 
The critical lines for four random particle noise realisations of the numerical cluster are over plotted with coloured lines.
The blue line is for the original N-body cluster E from the Level-2 Phoenix simulations. 
The discreteness of the particles in the simulation is especially important in the strongly magnified and highly nonlinear regime of the critical lines. 
Figure \ref{fig:noise_magnification} demonstrates the magnitude of the effect that significantly shapes the critical lines, even for this high-resolution simulation.
Multiple imaging caused by strong lensing also takes place in this highly magnified region. 
It is therefore important to understand and quantify the noise on the deflection angles and the highly magnified images. 
We describe the properties of the particle noise on the deflection angles in the next section.
\subsection{Deflection Angles}
\label{sec:Deflection_Angles}
Understanding the noise on the deflection angles is essential to quantify and describe the noise on the lensed images. 
We calculate the noise on the deflection angles by projecting all particles with a modified kernel, 
\begin{eqnarray}
\label{eq:sigma_alpha}
\bm W_{\sigma^2_{\bm\alpha}} \xx = \frac{c^2}{r_p^4} \left[ \underbrace{ G(r_p,l_p)^2  - 2 G(r_p,l_p) }_{\mathrm{A}} + \underbrace{1}_{\mathrm{B}}\right]
\label{eq:kernel_alpha}
\begin{pmatrix}
x^2 \\ y^2
\end{pmatrix}.
\end{eqnarray}
Here $\bm W_{\sigma^2_\alpha} \xx$ is a vector for the noise on the deflection angles in the x- and y directions, 
$c = m_p /(\pi \Sigma_c)$, $r_p$ is the distance of particle $p$ to the 
point $\bm x$, $l_p$ is the smoothing length and $G(r_p,l_p) = \exp(\frac{-r_p^2}{2 l_p^2})$.
The first two terms in $\mathrm{A}$ can be easily calculated by direct numerical summation. 
This is necessary since the smoothing length $l_p$ is different at each point. 
For the summation it is sufficient to consider points with $r_p \lessapprox 5 l_p$, since at larger distances for
both terms in A, the Gaussian kernel vanishes, $G(r_p,l_p) \rightarrow 0$. 
The second term, $\mathrm{B}$, is easily evaluated by convolving the grid via FFT. 
\begin{figure}
\begin{center}
\includegraphics[width=0.58\columnwidth]{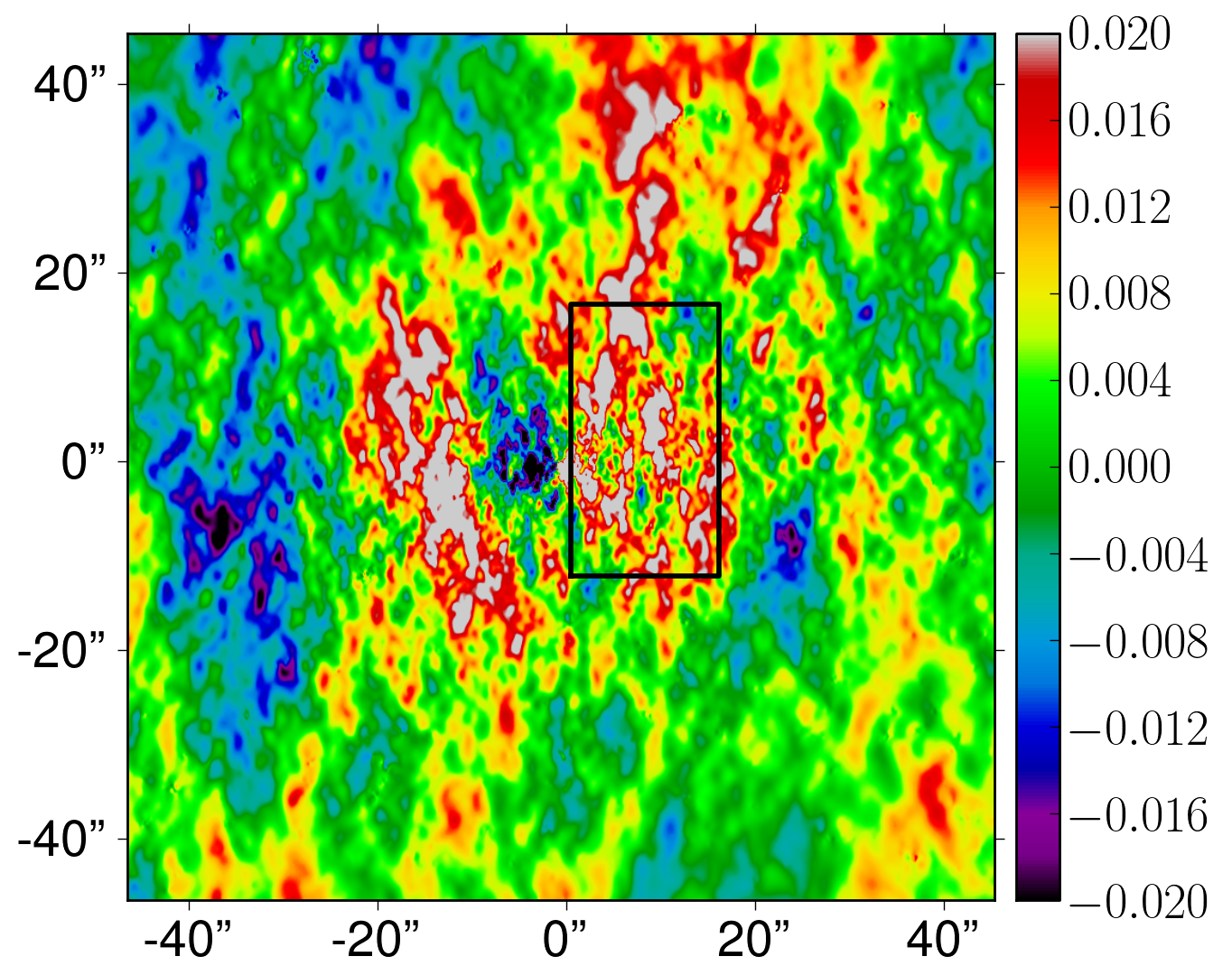}
\includegraphics[width=0.4\columnwidth]{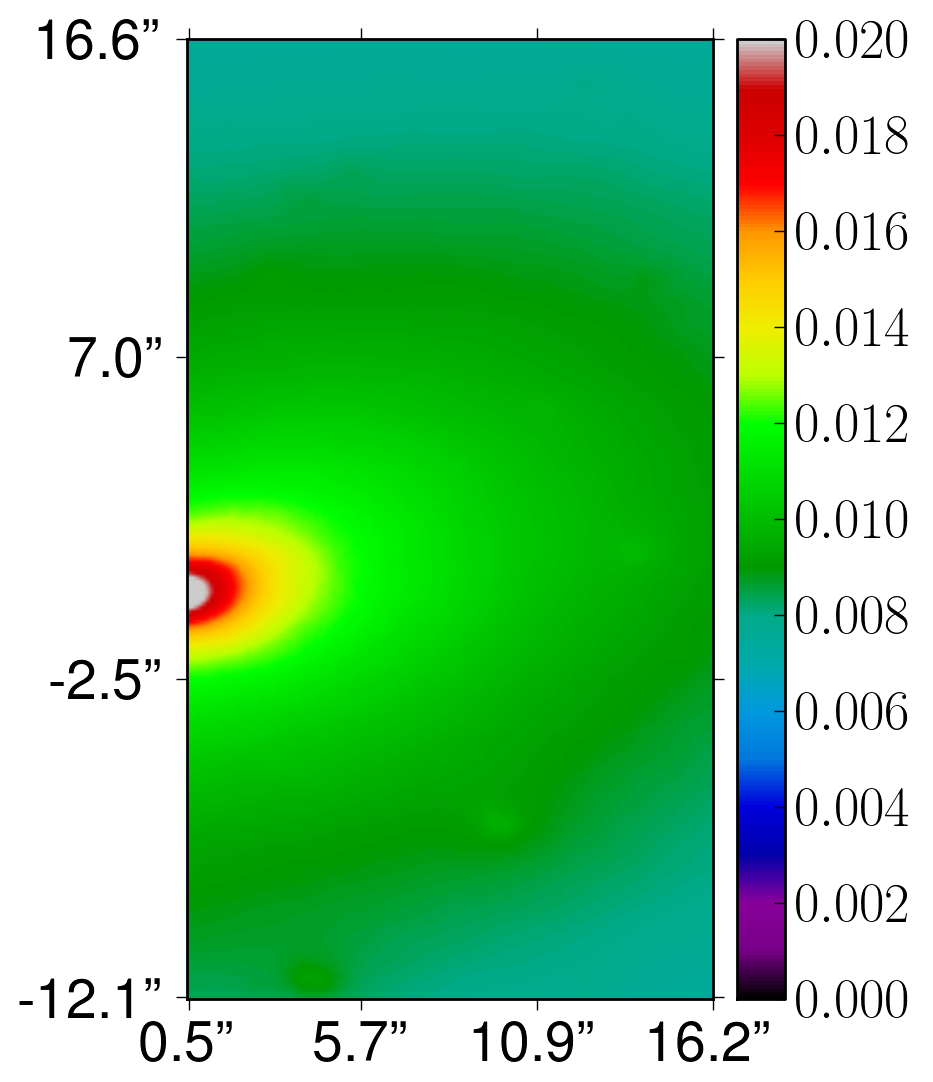}
\end{center}
\caption{ Difference between two random realisations of the particle noise of the deflection angles in x direction, $\alpha^1_x - \alpha^2_x$ (left panel). 
The marked region shows the size of a typical image from Sec.~\ref{sec:Images}. 
Standard deviation of the deflection angles in the marked region calculated with Eq.~\eqref{eq:sigma_alpha} (right panel).
For both panels the colour scale is in arcseconds.
}
\label{fig:noise_alpha}
\end{figure}

The noise on the deflection angles in x direction is shown in the right panel of Fig.~\ref{fig:noise_alpha} 
for the Level-2 version of cluster E. 
The side length of the figure is comparable to the size of a strongly lensed image.
Over the size of a typical giant arc, the effect of shot noise on the deflection angles is a smooth function.
The left panel of Fig.~\ref{fig:noise_alpha} shows a realisation of the particle noise on the deflection angles in the x direction. 
The marked region corresponds to the size of a typical multiply lensed image. Due to the $1/r$ dependence of the deflection angles, 
the particle noise is strongly correlated on long scales in the left panel of Fig.~\ref{fig:noise_alpha}.
These long-scale correlations will have a big effect on the calculation and the understanding 
of the noise on the lensed images presented in the next section.

\subsection{Lensed Images of a Gaussian Source}
\label{sec:Images}
\begin{figure}
\begin{center}
\includegraphics[width=0.48\columnwidth]{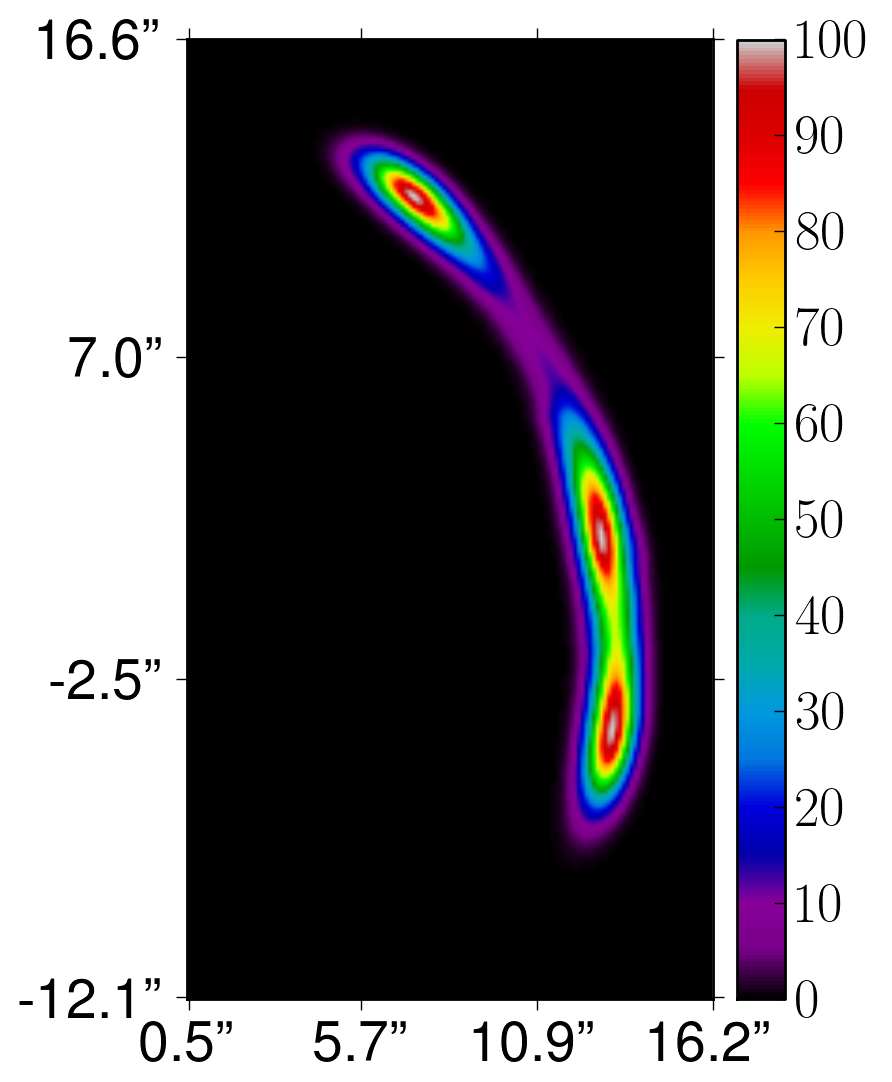}
\includegraphics[width=0.48\columnwidth]{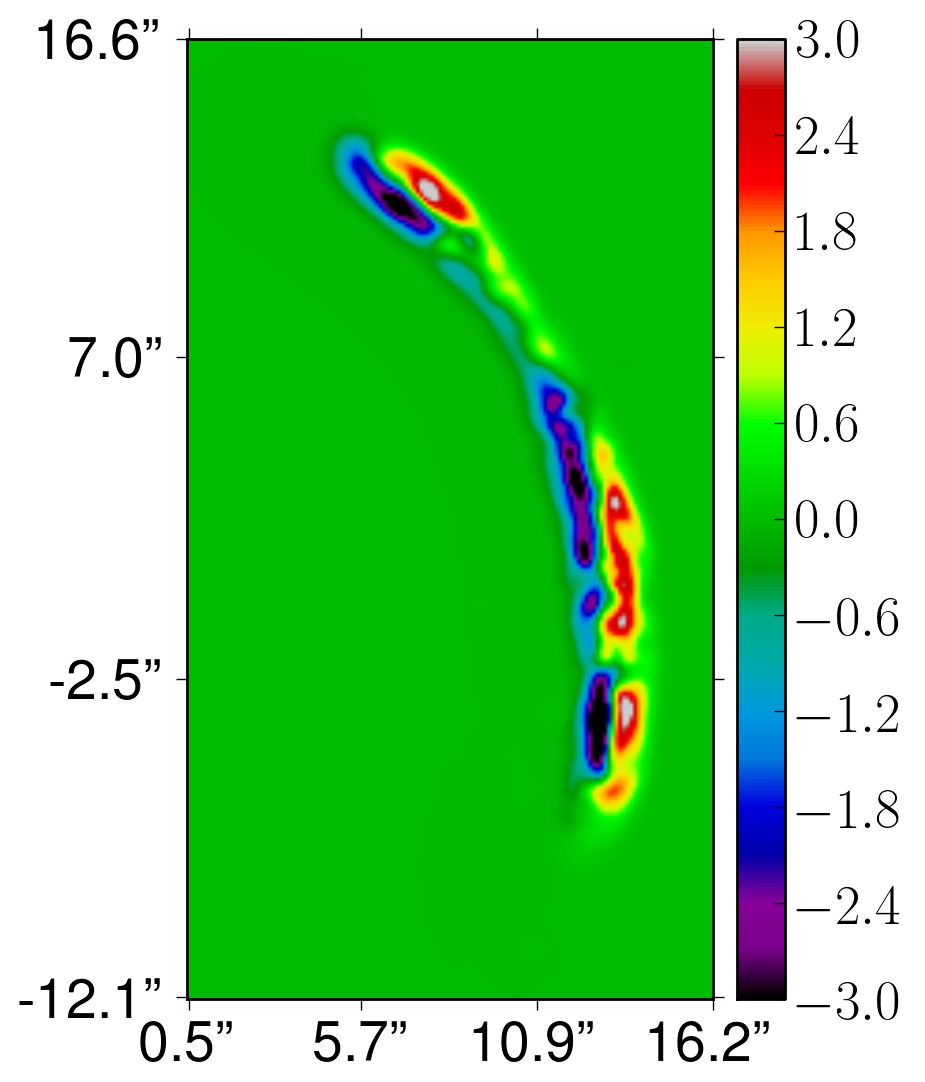}
\includegraphics[width=0.48\columnwidth]{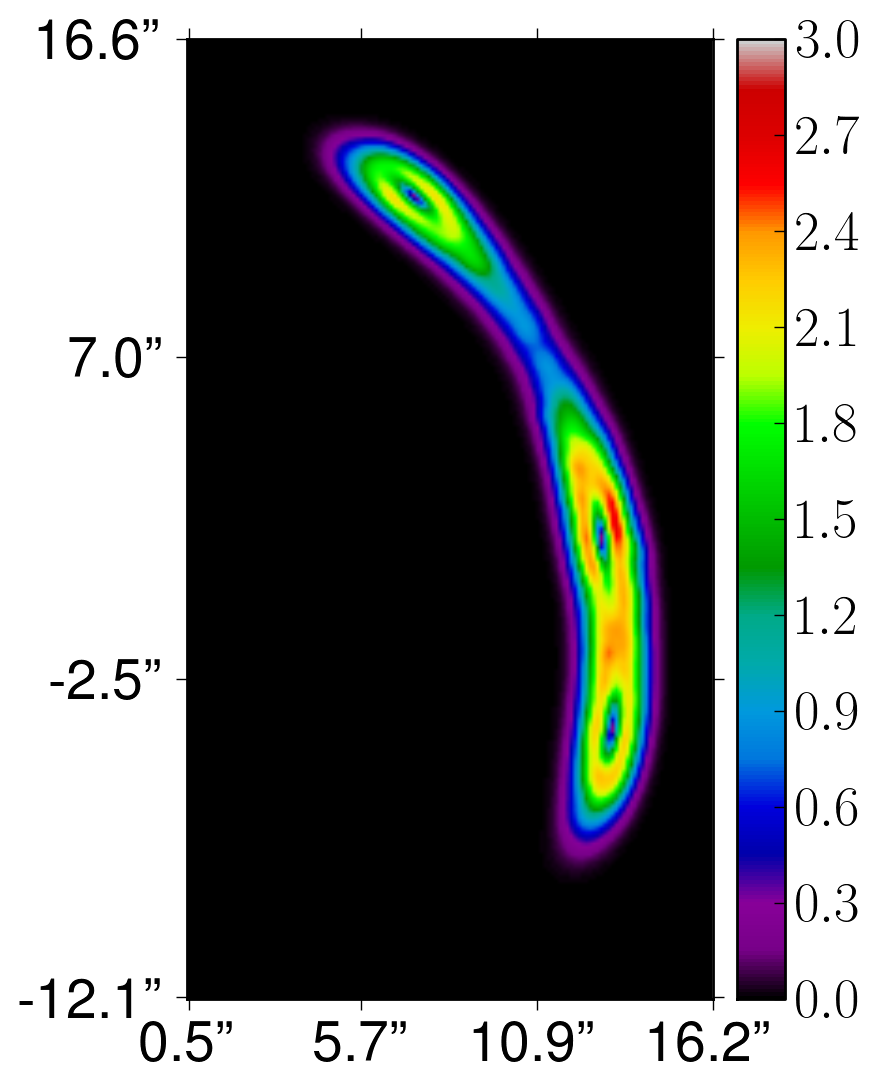}
\includegraphics[width=0.48\columnwidth]{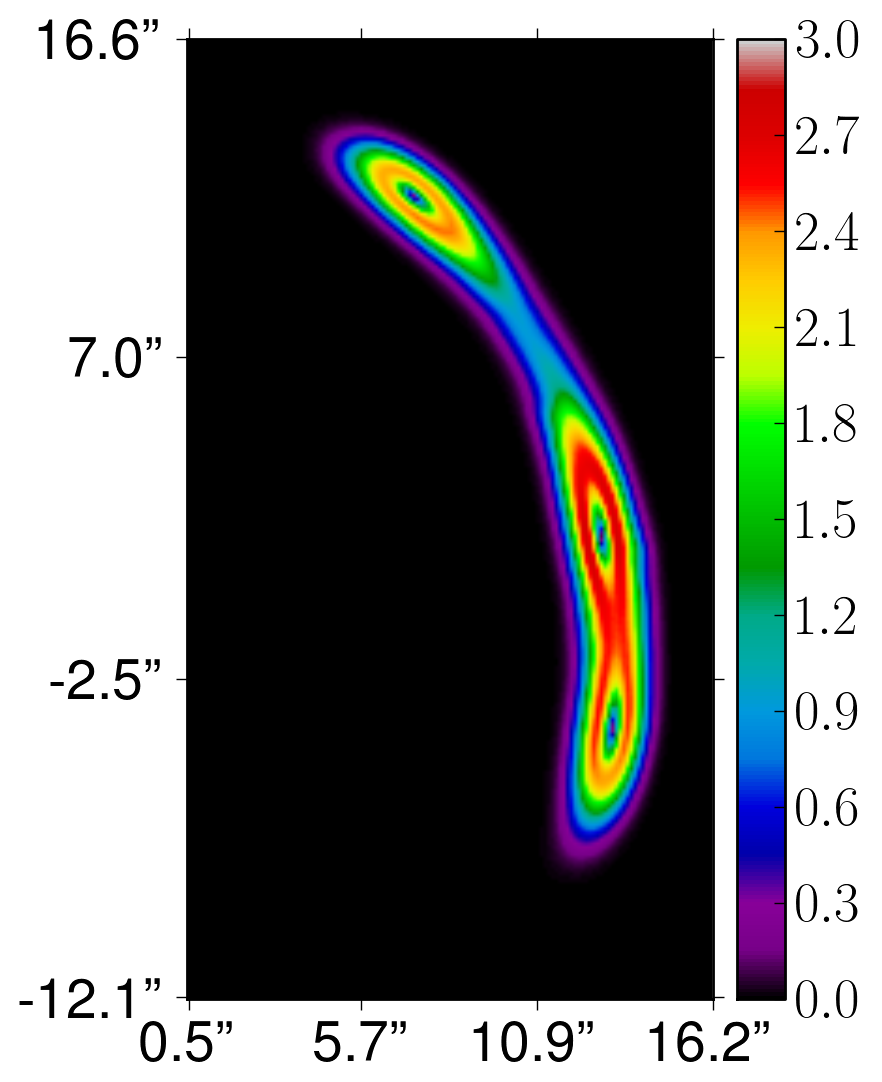}
\end{center}
\caption{
Image of a Gaussian source (top left) using the N-body cluster, brightness difference of two bootstrapped realisations (top right), 
standard deviation of 100 resamplings (bottom left) and analytic approximation using the noise of the deflection angles and a linearization of the lensing equation (bottom right).
}
\label{fig:images_constant_source}
\end{figure}
In order to understand the effect of particle noise on the strongly gravitationally lensed images of a background source, 
we simulate a gravitational lens system by placing a source galaxy at $z=2.0$ 
behind cluster E at Level-2 resolution, $(4.4'',4.1'')$ away from the cluster centre in projection. 
The source has a Gaussian surface brightness distribution with a FWHM size of $0.51''$ and a peak surface brightness of 100. 
The resulting gravitationally lensed images are shown in the top left panel of Fig.~\ref{fig:images_constant_source}. 
Multiple images of a background source form in regions of high magnification, where the effect of the particle noise, due to the  discreteness of the N-body simulations, is the largest. 
It is very important, therefore, to understand how the noise on the cluster surface mass density and deflection angles is propagated to the lensed images. 
In the following, we quantify this noise from the bootstrapped re-samplings of the N-body particles (compare with Sec.~\ref{sec:Particle_noise}), 
as well as from analytical approximations.

For each bootstrapped particle distribution $i$ of the cluster lens, 
we calculate the deflection angles $\bm\alpha_i$ and use them to lens an identical source surface brightness distribution 
$\bm s$ for each bootstrap resampling of the lens. 
The corresponding images, $\bm d_i$, differ from each other because of the particle noise.

The lower left panel of Fig.~\ref{fig:images_constant_source} shows the standard deviation of 100 lensed images of the same background source calculated in
the same way from 100 different bootstrapped cluster resamplings; the noise on the image brightness is as large as $\sim 10 \%$ of the image surface brightness. 
The origin of the image brightness fluctuations due to the particle noise and their distribution can be understood qualitatively by
considering the lens equation $\bm y = \bm x - \bm \alpha\xx$. According to this equation, 
each point $\bm x$ on the lens plane is deflected by a deflection angles $\bm \alpha$ to a point $\bm y$ on the source plane. 
As shown in Sec.~\ref{sec:Deflection_Angles}, the noise associated with the discrete particle distribution is responsible 
for local fluctuations on the gravitational potential $\psi\xx$ of the lensing cluster and therefore on the deflection angles $\bm \alpha=\nabla \psi\xx$. 
The noise on the lensed images is then just the effect of the particle noise on the cluster surface mass density propagated to the images via the deflection angles. 
From the left bottom panel of Fig.~\ref{fig:images_constant_source}, 
it is clear that a large number of realisations is needed in order to reduce the statistical error of the bootstrap noise estimation. 
More quantitatively, we can extend the analytic noise analysis of the deflection angles in Sec.~\ref{sec:Deflection_Angles} to the noise on the simulated images, 
by  linearizing the lens equation and approximating the difference in the surface brightness between two images $\delta \bm d$ by (\cite{Koopmans_2005}),
\begin{equation}
\delta \bm d \approx - (\delta \bm\alpha_x \, \partial_x \bm s + \delta \bm\alpha_y \, \partial_y \bm s).
\label{eq:lin_delta_d}
\end{equation}
For a given source surface brightness distribution, fluctuations in the image surface brightness are a linear combination of the fluctuations in the deflection angles.  
The bottom right panel of Fig.~\ref{fig:images_constant_source} shows this approximation for the noise on the images using the noise on the deflection angles. 
If we compare the two lower panels of Fig.~\ref{fig:images_constant_source}, we see that this linearization overestimates the noise where the noise is high.
From the above equation, we also expect the noise to follow the distribution of the gradient of the source surface brightness. 
In particular, small changes in the deflection angles can be strongly amplified if the gradient of the source surface brightness distribution is large. 
In Sec.~\ref{sec:images_realistic}, we discuss the effect of particle noise for a source surface brightness distribution with varying level of smoothness.

The top right panel of Fig.~\ref{fig:images_constant_source} shows the difference in brightness 
$\bm d_0 - \bm d_1$ for each pixel of the image plane, as an example, for two particular cluster re-samplings $i=0$ and $i=1$. 
The observed surface brightness difference can be attributed mainly to two effects: 
A global shift to the left by $< 0.1\arcsec$ of $\bm d_1$ relative to $\bm d_0$, and smaller scale differences. 
We start by discussing the origin of this global shift. 
We will see in the following that this shift is equivalent to an unobservable shift of the caustic relative to the source.
We will therefore present a method to separate and subtract the contributions of this shift to the noise on the image brightness later on in this section.

\subsubsection{Global shift of the lensed images}
\label{subsubsec:global_shift_images}
Although, small scale fluctuations of the lensed images can be significant, 
an overall shift in the image position also contributes significantly to the difference between two bootstrapped images.
In the following, we shall discuss the differences in the deflection angles and the caustic structure between two re-samplings in more detail. 
This will help us to eliminate most of the brightness difference in the top right panel of Fig.~\ref{fig:images_constant_source} by a relative source - caustic shift, 
and will provide a more physical and observationally motivated measure for the noise on numerically lensed images.

\begin{figure}
\begin{center}
\includegraphics[width=1.0\columnwidth]{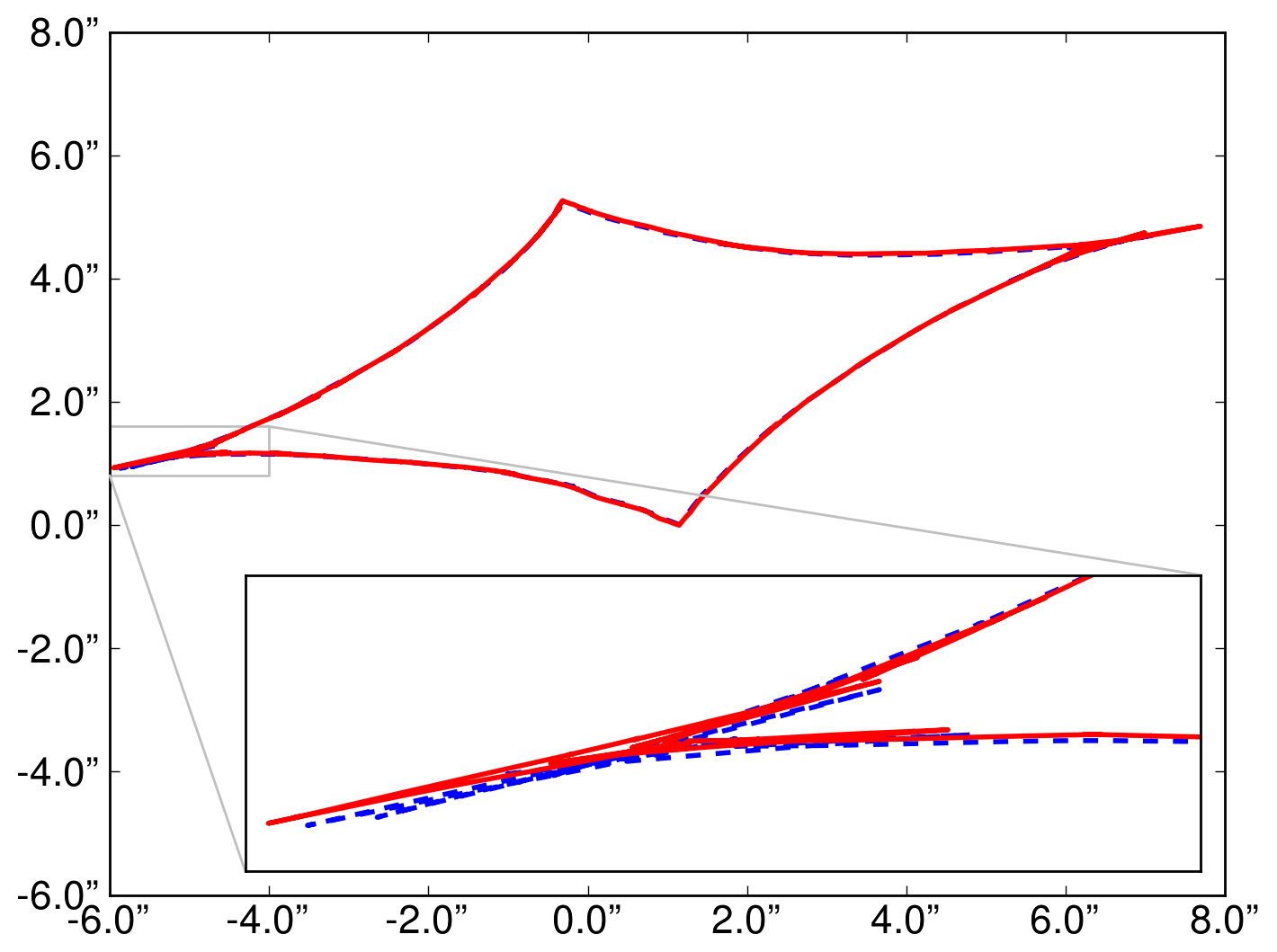}
\end{center}
\caption{Caustics of two different particle noise realisations. 
Due to long-scale correlations of the deflection angles (cf. Fig.~\ref{fig:noise_alpha}) and the finite size of the image, 
random particle noise will result in a shift of the caustic. 
}
\label{fig:noise_caustic}
\end{figure}

In the left panel of Fig.~\ref{fig:noise_alpha}, the difference between two deflection angle maps calculated from two bootstrapped clusters is shown.
The size of a typical image, as for example in the top left panel in Fig.~\ref{fig:images_constant_source} is marked as a rectangle.
Because of the finite size of the area covered by an image on the lens plane and because of the long scale noise correlations on $\bm \alpha$, 
the mean of the noise realisation of the deflection angles within the area covered by the image is positive. 
This positive deviation of the deflection angles, also visible as a global shift of the caustic structure (see Fig.~\ref{fig:noise_caustic}), 
will then result in a mean shift of the lensed images, or equivalently of the source on the source plane.
Therefore, while a mean constant additional deflection angle does not change any of the physical parameters of the lens, 
it changes the relative position of the caustic and the (arbitrarily fixed) source.

\subsubsection{Small scale surface brightness fluctuations}
\label{subsubsec:small_scale_brightness_fluctuations}
Any shift of the caustic position can be compensated by shifting the source position accordingly.
Therefore, a simple shift in the source position eliminates most of the image brightness difference in the top right panel of Fig.~\ref{fig:images_constant_source}.
In the previous section we therefore made an assumption. 
We used an identical source for the shifted and the not-shifted caustic structure and we therefore compared two non-equivalent images.

Following a similar chain of argument, it is possible that the re-sampled cluster deflection angles will differ in higher order derivatives of the deflection angles as well, 
such as magnification, shear or flexion or more likely, a combination of all of those. 
Since none of the sources intrinsic properties such as position, size or morphology are known, 
we will assume in the following as little as possible about the source.
The method described below is closer to an observational point of view and only constrains the source by its regularity. 
We therefore rephrase the problem of the comparison of two equivalent images as follows.
Two deflection angle maps are given and the first of the two images is fixed as reference image.
What is the most similar image we can find using the second deflection angle map, 
if we are only allowed to use a relatively regular source?

This question can be more easily answered within a Bayesian formulation of gravitational lensing.
We try to find the closest image $\bm d_1$, to a reference image $\bm d_0$ under some regularity conditions. 
For a given, fixed smoothness of the source it is in general not possible to perfectly recreate the input image $\bm d_0$ exactly  
using the deflection angles $\bm \alpha_1$ and a smooth source $\bm s_1$. 
It is, however, possible to reconstruct a close image $\bm d_1$.
We will try to find the best reconstructed image $\bm d_1$, which is as close as possible to the input image $\bm d_0$, 
and at the same time keeping its source $\bm s_1$ regular.
Our source reconstruction is pixelized on a non-regular triangulation and the smoothness is only constrained
by a curvature regularisation (\cite{Vegetti_2009a}). 
For a given input image $\bm d_0$ we maximize the Bayesian posterior 
\begin{equation}
 \log P(\bm s_1|\bm d_0, \bm\alpha_1,\lambda, \bm H_s) = \chi^2 + \lambda_s^2 || \bm H_s \bm s_1 ||^2_2
\end{equation}
where $\chi^2$ is between both images, $\bm H_s$ is the form and $\lambda_s$ the strength of the regularisation of the source. 
We assume a quadratic Gaussian prior for the regularisation which is centred at $\bm s = 0$. The regularisation strength $\lambda_s$ is found 
self consistently from the data themselves by maximising the posterior 
\begin{eqnarray}
P(\lambda_s|\bm d_0, \bm\alpha_1,\bm H_s) =  \frac{P(\bm d_0|\lambda_s,\bm\alpha_1,\bm H_s)P(\lambda_s)}{P(\bm d_0|\bm\alpha_1,\bm H_s)} \nonumber \\
 \propto \int \! \mathrm{d}\bm s_1  \, P(\bm d_0 |\bm s_1, \lambda_s, \bm\alpha_1, \bm H_s) P(\bm s_1|\bm\alpha_1, \bm H_s).
\end{eqnarray}
Here we assume a prior $P(\lambda_s)$ which is flat in $\log \lambda_s$.
This method uses minimal assumptions about the source to find the best matching image for an input image. 
\begin{figure}
\begin{center}
\includegraphics[width=0.48\columnwidth]{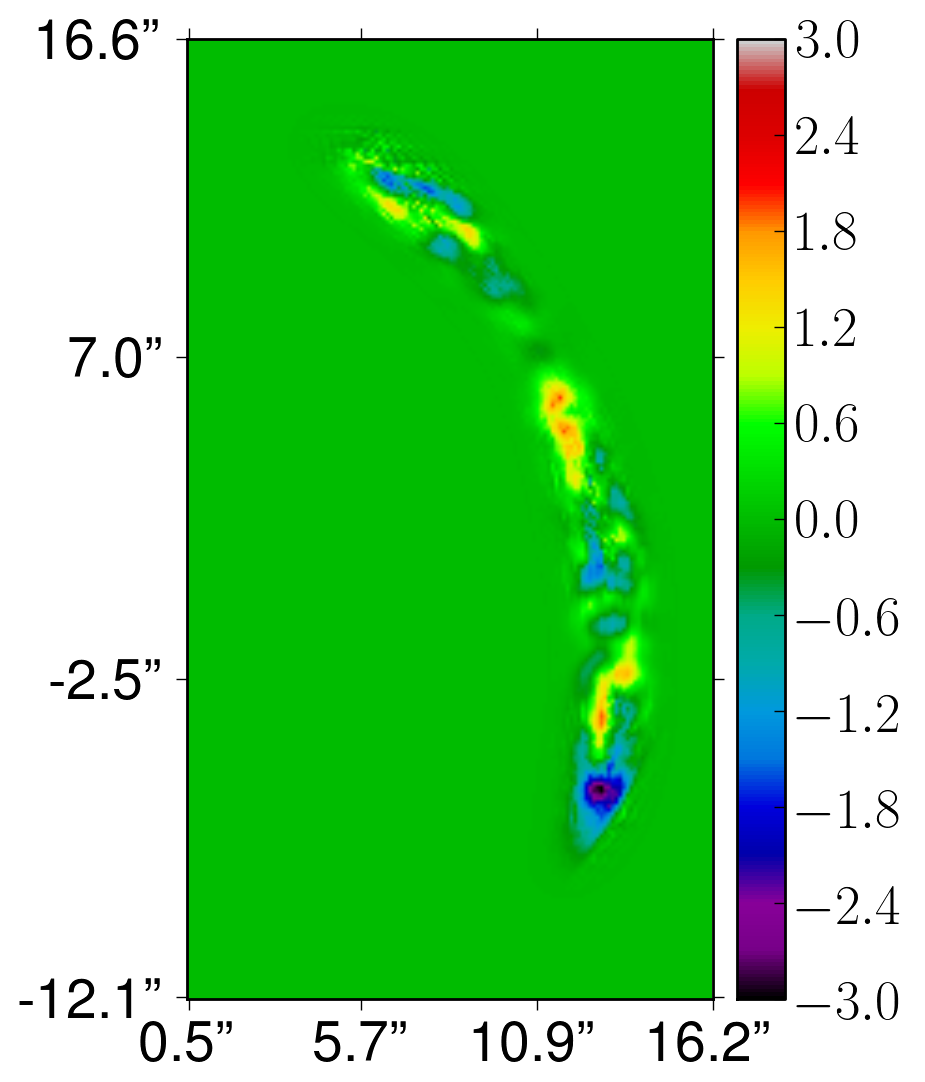}
\includegraphics[width=0.48\columnwidth]{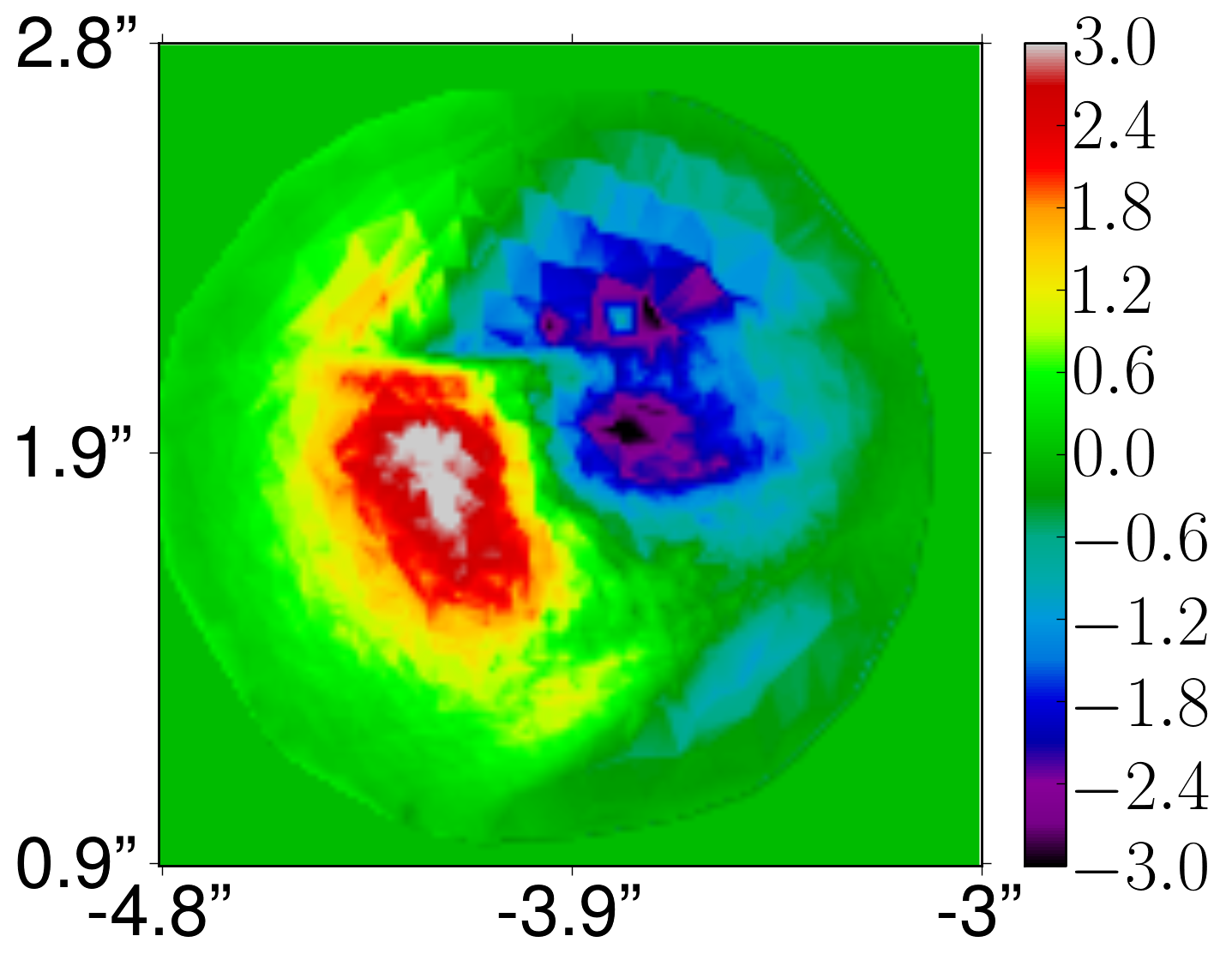}
\includegraphics[width=0.48\columnwidth]{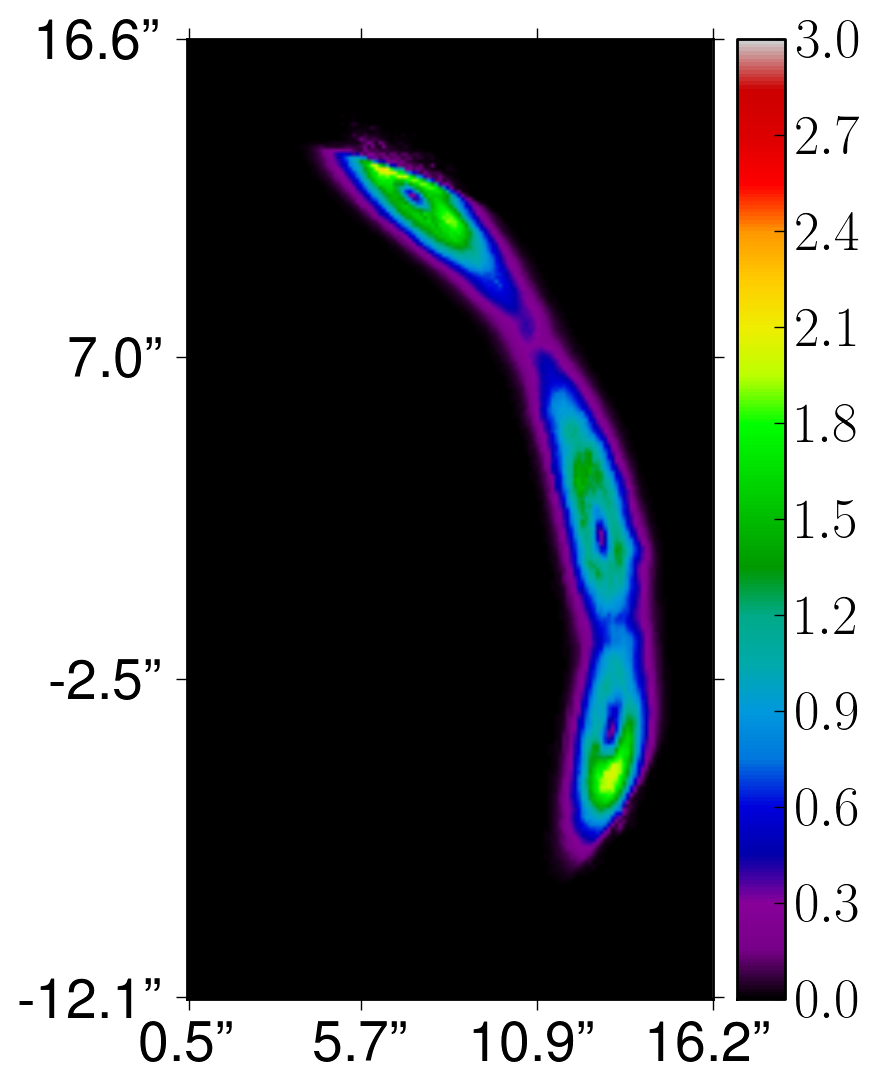}
\includegraphics[width=0.48\columnwidth]{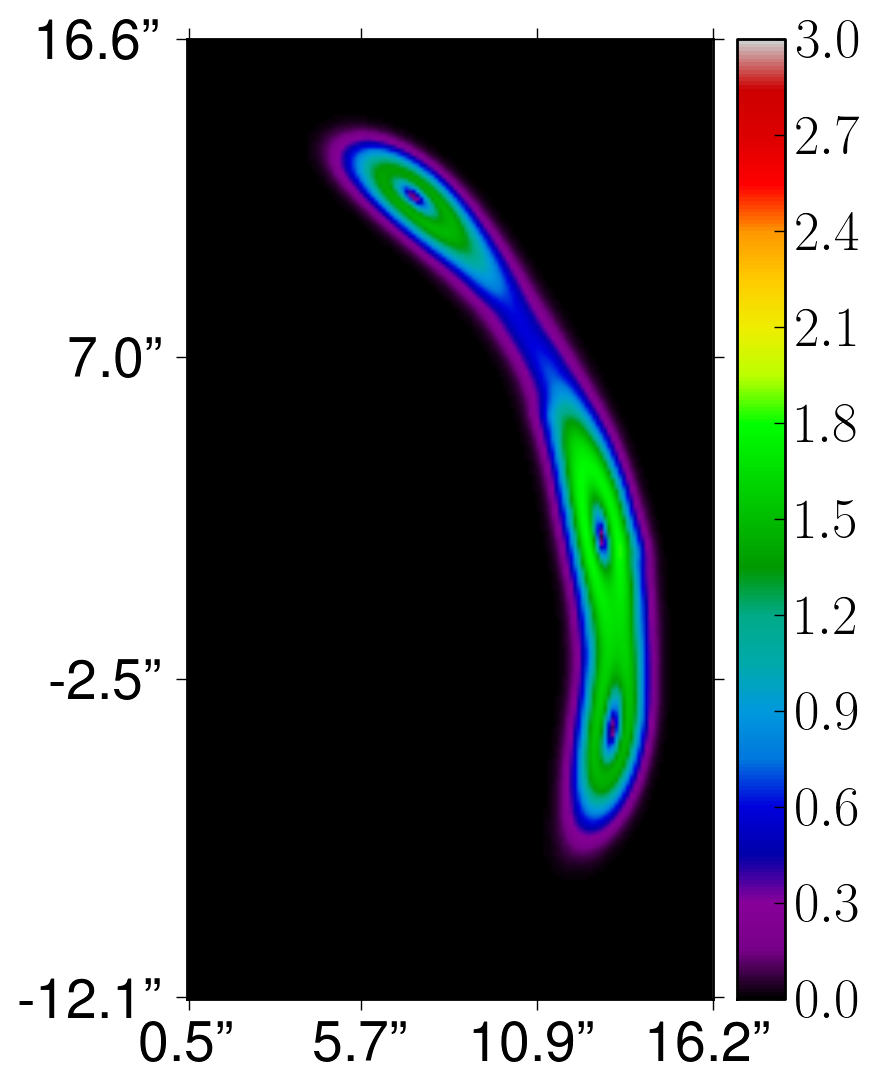}
\end{center}
\caption{
Difference in the image brightness using a Bayesian source reconstruction (top left panel), colorscale identical to Fig.~\ref{fig:images_constant_source} top right panel. 
Difference between original source and reconstructed source shows that the biggest effect is a shift in position (top right panel). 
Standard deviation of 100 bootstrapped realisations of the N-body cluster, for each cluster the best image matching an input image is found (bottom left panel). 
Standard deviation approximation with analytic formula using $\sigma_\alpha$ and a linearization of the lensing equation (bottom right panel).
\label{fig:images_bayes_fourier}
}
\end{figure}
The pixelized brightness difference $\bm d_0\xx - \bm d_1\xx$ of the reconstruction is shown in the top left panel of Fig.~\ref{fig:images_bayes_fourier} for the same example as in the top right panel of Fig.~\ref{fig:images_constant_source}.
This method automatically corrects for the shift in the deflection angles or in the caustic in Fig.~\ref{fig:noise_caustic} and compares images produced by equivalent sources. 
The surface brightness difference between the two sources is shown in the top right panel of Fig.~\ref{fig:images_bayes_fourier}.
As explained above, the reconstructed source is automatically shifted by the algorithm in order to 
find an equivalent source which now reproduces the reference image as well as possible.
By applying this method to $N_\mathrm{b} = 100$ bootstrapped cluster resamplings we simulate the particle noise on the 
pixelized image brightness distribution, $\bm\sigma^2_{\bm d} \xx = 1/(N_\mathrm{b}-1) \sum^{N_\mathrm{b}}_{i=1}\left(\bm d_0 - \bm d_i\right)^2$. 
The noise is shown in the bottom left panel of Fig.~\ref{fig:images_bayes_fourier}.
By allowing the algorithm to automatically adapt the source, the noise on the images is reduced by a factor of two.
In other words, half of the noise on the image brightness in the lower left panel of Fig.~\ref{fig:images_constant_source}
can be attributed to a simple relative shift of the source to the caustic between different bootstrap realisations.
The sharp cuts at the top and the bottom of the arc in the lower left panel of Fig.~\ref{fig:images_bayes_fourier} are due to the caustic structure.
The single imaged regions of the source are less noisy by a factor of $\sim 100$ than the threefold imaged parts. 
The reconstruction algorithm for the source can more easily fit the single-imaged regions than the threefold imaged parts of the brightness distribution.
Therefore, the noise is not visible in these regions in Fig.~\ref{fig:images_bayes_fourier}. 
If we assume that the Bayesian source reconstruction mainly corrects for the source shift, we also can approximate this noise analytically. 
Shifting the source keeps the gradient of the source $\nabla \bm s$ constant on the image plane. But we have to correct the deflection angles by the mean source shift, effectively calculating
$\delta \bm d \approx - (\delta \bm \alpha - \left\langle\delta \bm \alpha\right\rangle) \nabla \bm s$ where the mean $\left\langle\delta \bm \alpha\right\rangle$
is evaluated over the size of the images. The bottom right panel of Fig.~\ref{fig:images_bayes_fourier} shows the corrected version of the linear noise approximation for the image brightness. 
Even though this is a very simplified and fast approximation to the noise, the result is very close to the accurately simulated noise.

\subsection{Lensed Images of a Realistic Source}
\label{sec:images_realistic}

\begin{figure}
\begin{center}
\includegraphics[width=0.48\columnwidth]{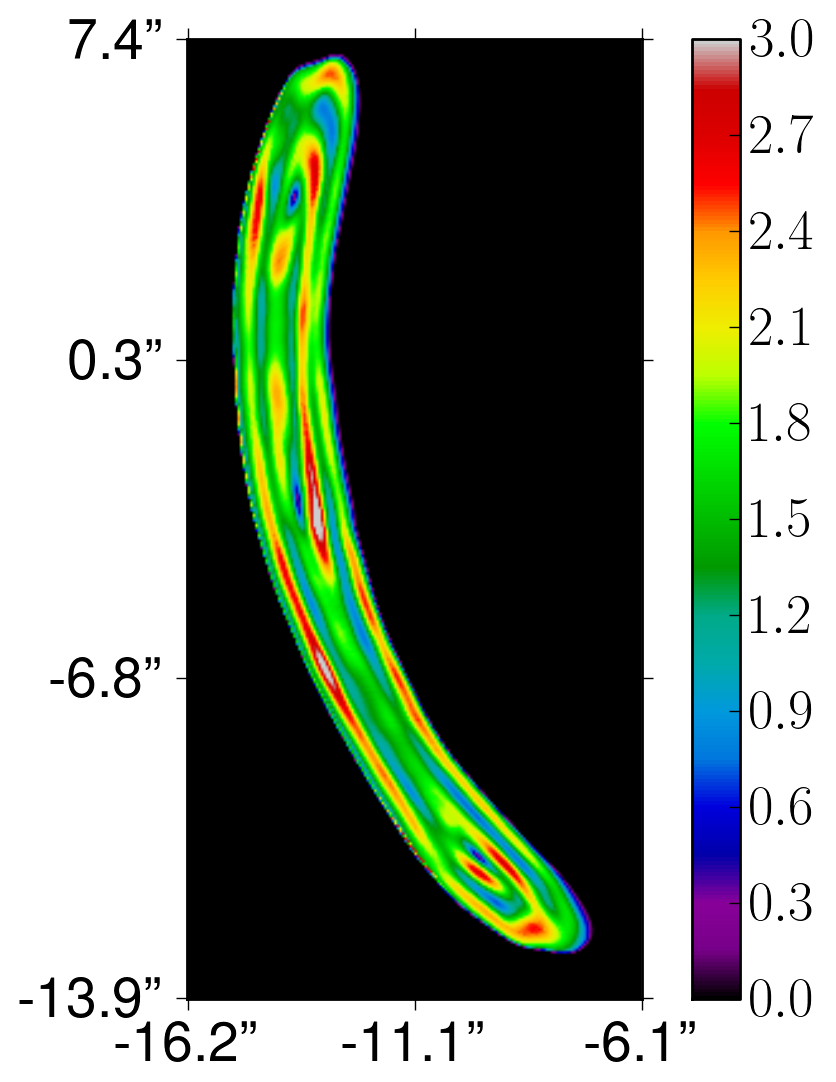}
\includegraphics[width=0.48\columnwidth]{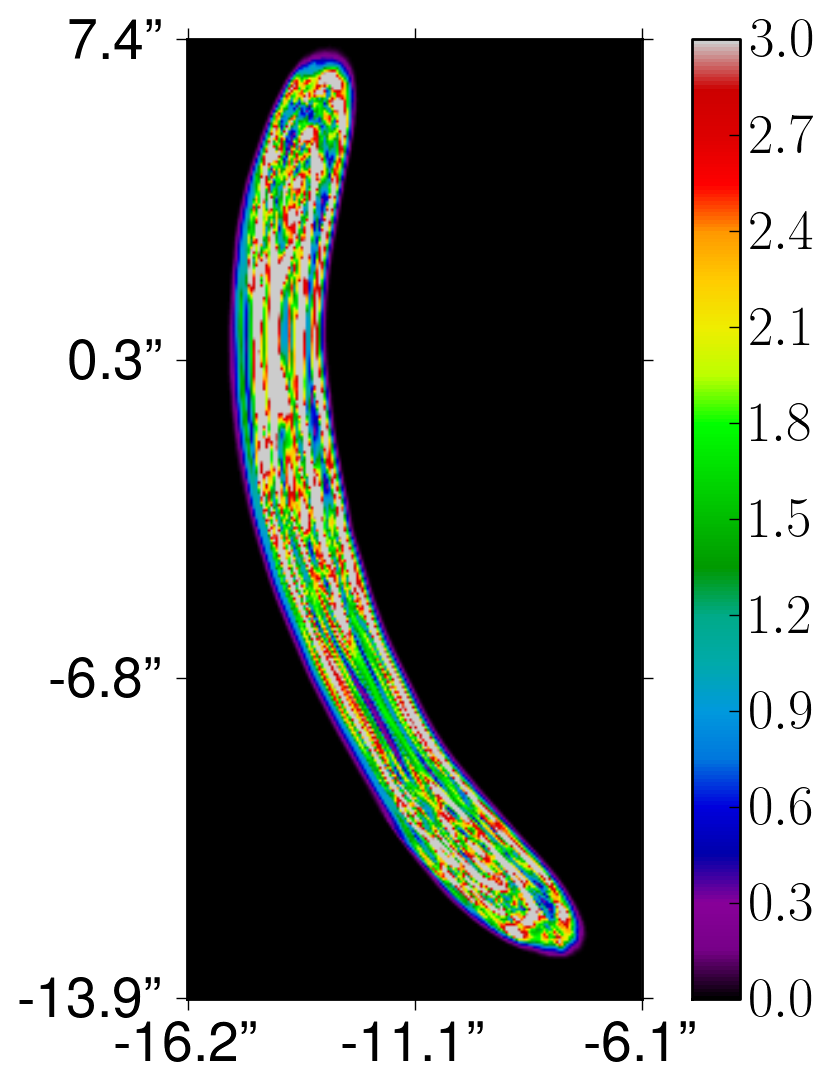}
\includegraphics[width=0.48\columnwidth]{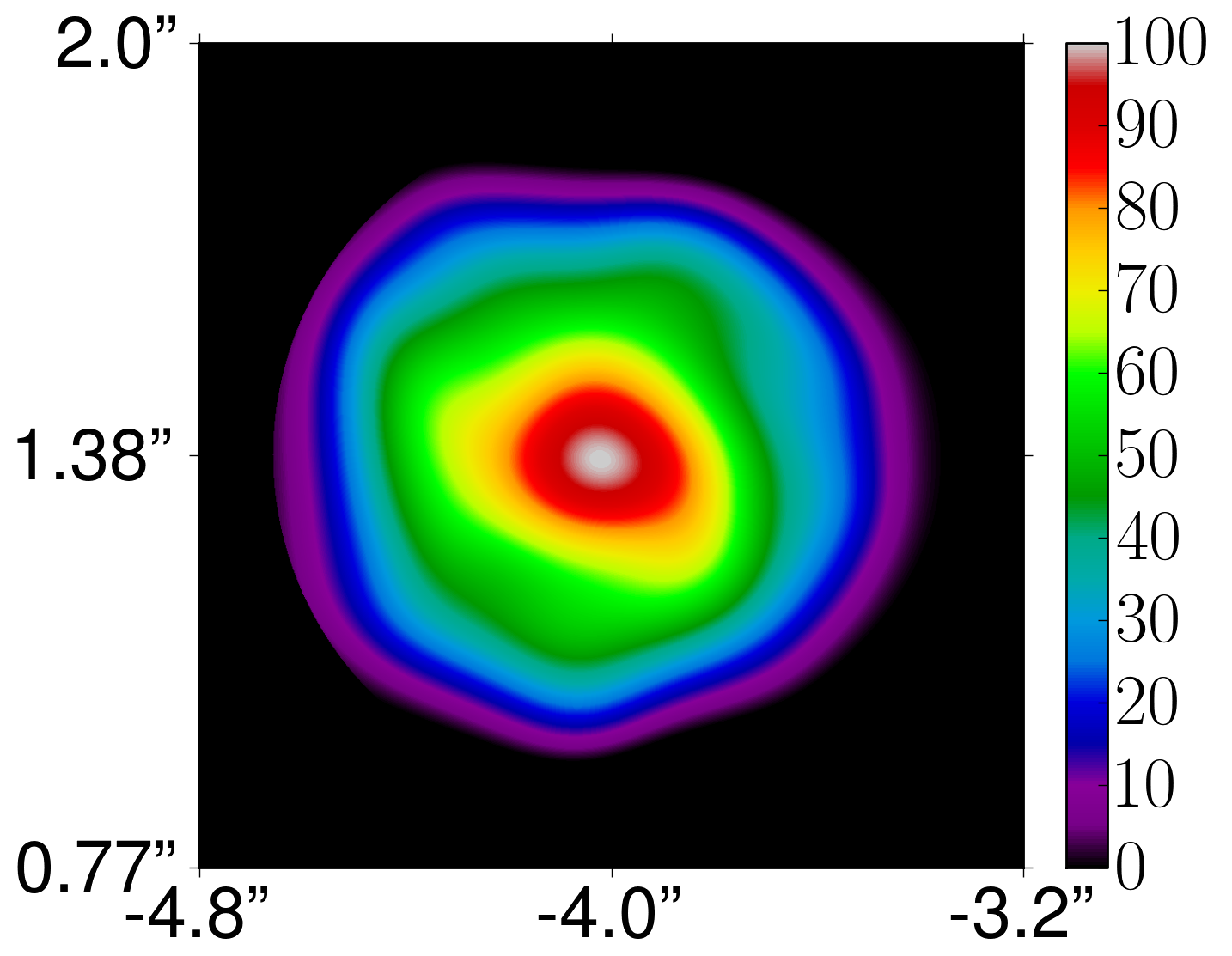}
\includegraphics[width=0.48\columnwidth]{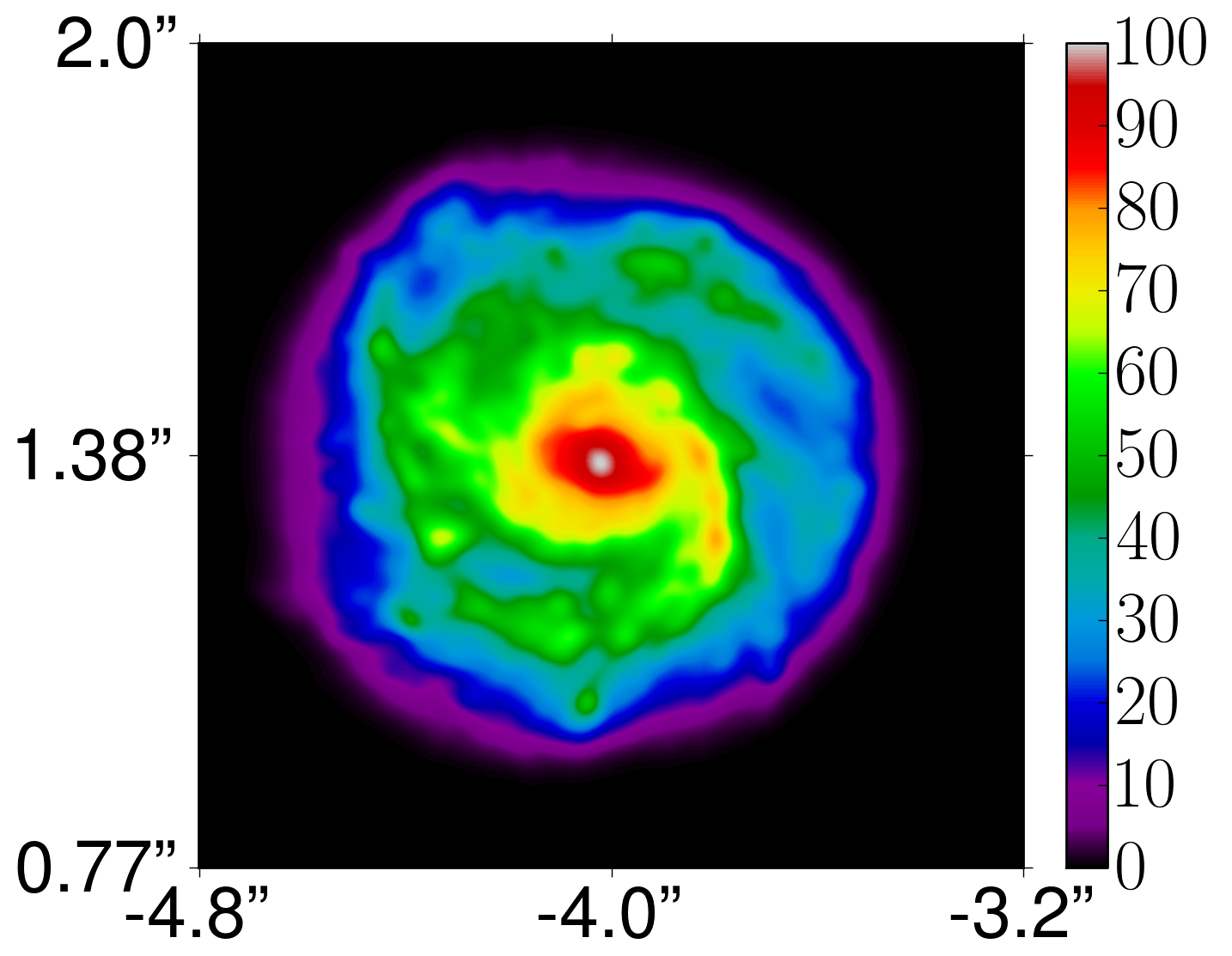}
\end{center}
\caption{
Same as the bottom left panel of Fig.~\ref{fig:images_constant_source} but for a different source position and the source brightness distributions in the bottom panels. 
An increased structure of the source and therefore an increased source gradient increases the noise on the image brightness. 
}
\label{fig:images_noise_different_sources}
\end{figure}

The noise on the surface brightness distribution of the lensed images only weakly depends on the image position, but it strongly depends on the source gradient.
This can be understood from the analytical approximation in Eq.~\ref{eq:lin_delta_d}, since the noise on the deflection angles 
in the top right panel of Fig.~\ref{fig:noise_alpha} is very smooth in the region where strongly magnified images occur at $x \approx \pm 15\arcsec$. 
As an example we show the different noise maps of the image brightness in the top panels of Fig.~\ref{fig:images_noise_different_sources} 
for two more structured sources. The respective sources are shown in the lower panels. 
Both sources are located at $(-4\arcsec,1.38\arcsec)$, the FWHM size is the same as the Gaussian source in Fig.~\ref{fig:images_constant_source}, $0.51\arcsec$, 
and the maximum brightness is scaled to 100.
The source on the left is a smooth version of the right source. 
The magnitude of the noise in the top left panel is similar to the noise on the image for a Gaussian source. 
This can be seen by comparing Fig.~\ref{fig:images_noise_different_sources} to the lower left panel of Fig.~\ref{fig:images_constant_source}.
The distribution of the noise follows that of the source gradient on the lens plane. 
The more irregular source in the right panel also shows a significantly increased noise. This dependence is easily understood from the 
linearization in Eq.~\ref{eq:lin_delta_d}.\\

\section{Noise Scaling} 
\label{sec:scaling}
In the previous sections, we have calculated the particle noise due to the discreteness of the N-body simulation on different lensing properties. 
The results were calculated for a simulation with $\Npart$ particles and a 3D adaptive smoothing algorithm that 
adapts the smoothing length for each particle to the distance of its $\Nngb$-th neighbour in 3D. 
In this section, we will derive how the particle noise in the simulation of gravitational lensing 
changes as a function of particle number $\Npart$ and the number of smoothing neighbours $\Nngb$. 
Increasing the number of particles inside the smoothing kernel, $\Nngb$, will result in a more smoothed mass distribution and therefore reduced noise
on all of the lensing properties. It will, however, also smooth out the physical substructures that were part of the simulation. 
Increasing the particle number while keeping the number of neighbours constant will increase the mass resolution of the simulation and therefore also reduce the noise,
at the cost, however, of the computational load.

In principle, we could numerically simulate a change in $\Npart$ and $\Nngb$ by resampling the mass density and then re-calculate the smoothing length for each particle
and all of the lensing properties for each of the re-projections of the 3D particle distributions.
In practice, we develop a method that allows us to convert the 3D adaptive smoothing lengths into 2D smoothing lengths and derive analytic expressions 
for the scaling of the noise in 2D, while preserving the information of the 3D particle density distribution.

\subsection{Smoothing in 2D}
\label{subsec:2D_Approximation}
Until now we calculated the noise on all of the lensing properties from the 3D particle distribution and we used a smoothing length that was adaptive with the 3D particle density distribution. 
We will now introduce an approximation that allows us to keep most of the 3D information, while performing the noise calculations in 2D. 
The 2D version obviously increases the speed of the calculations substantially, and it also allows us to derive the scaling properties of 
the noise on the surface mass density and the deflection angles with the particle number and the number of smoothing neighbours.

By comparing Eqs.~\eqref{eq:sigma2_calc} and \eqref{eq:sigma}, we find that on average, on scales larger than the smoothing length $l_i$ of each particle, 
the contribution of each particle $i$ to the variance of the surface mass density, $\sigma^2_\Sigma$, and to the surface mass density, $\Sigma$, differs by a factor of 
\begin{equation}
 f_i \propto \frac{\int W(l_i) \mathrm{d}^2\bm x}{ \int W^2(l_i) \mathrm{d}^2\bm x} = \frac{9}{5 \pi l_i^2}.
\label{eq:w2factor}
\end{equation}
The constant factor $9/(5\pi)$ depends weakly on the form of the smoothing kernel. 
Here, we used the kernel $W \propto (1 - r^2/l^2)^2$ for $r \leq l$, for details and other kernels see also Appendix~\ref{sec:Variance_Smoothed_Particle_Distribution}.
Eq.~\eqref{eq:w2factor} states that smaller particles from high-density regions contribute proportionally with a factor of $1/l^2$ to the variance
of the surface mass density. 
We can use this information to derive a 2D approximation to the noise on the surface mass density.
To this end, we define an effective smoothing length, $l_{\mathrm{eff}} \xx$ in 2D, which is
a integrated average of all line of sight N-body particles,
\begin{equation}
 l_{\mathrm{eff}} \xx = \left(\frac{1}{N_z}\sum_{i=1}^{N_z} \frac{1}{l_i^2}\right)^{-1/2}
\label{eq:leff_def}
\end{equation}
This allows us to assign a single smoothing length to a particular 2D position on the lens plane,
and simultaneously taking into account the 3D density distribution.
Note that, in the limit of an extremely high-resolution grid on the lens plane and a finite N-body particle number,
not all positions will be occupied, and $\leff = l$.
With this simplification we are able to approximate the variance of the surface mass density of Eq.~\eqref{eq:sigma2_calc}, in terms of the 2D surface mass density $\Sigma$,
\begin{equation}
 \sigma^2_\Sigma\xx \approx \frac{9}{5\pi}\frac{m_p}{l^2_{\mathrm{eff}}\xx A}\Sigma\xx.
\label{eq:approx_sigmas_sigma}
\end{equation}

\subsection{Noise in 2D}
\label{subsec:Fast_Approximate_Method}
In this section, we introduce some useful expressions 	
for the noise on the lensing properties using the 2D smoothing approximation derived above. 
This will simplify the derivation in the next section and will allow us to calculate the noise on a high-resolution grid in a fast and easy way, 
by eliminating the need for a tree-based evaluation of the long-range terms of the deflection angles (e.g. the $1/r^2$ dependence of the variance).

We start by considering the definition of the effective smoothing length for each point on the lens plane $l_{\mathrm{eff}} \xx$ as presented in Eq.~\eqref{eq:leff_def}. 
This smoothing length defines the correlation length of the noise for each point in 2D.
All we need for the first order covariance matrix is the amplitude of the uncorrelated noise $\tilde\sigma_\Sigma$ at each point on the lens plane, which 
can be calculated from the correlated noise in Eq.~\eqref{eq:sigma2_calc}.
In other words, the correlation introduced by smoothing the original N-body particles with a smoothing kernel of size $\leff$
has decreased the uncorrelated noise to the expression given in Eq.~\eqref{eq:sigma2_calc}. 
By deconvolving each point on the lens plane with a smoothing kernel of size $l_{\mathrm{eff}}$, the correlated noise, $\sigma^2_\Sigma$, is increased again by
\begin{equation}
\tilde\sigma_\Sigma^2\xx = \frac{5 \pi l_{\mathrm{eff}}^2}{9} \sigma^2_{\Sigma}\xx ,
\label{eq:corrected_variance}
\end{equation}
as demonstrated in the Appendix \ref{sec:Variance_Smoothed_Particle_Distribution}.
In order to obtain the uncorrelated noise amplitude, $\tilde\sigma_\Sigma$, we therefore have to increase the noise from Eq.~\eqref{eq:sigma2_calc}.

These considerations allows us to simulate the effect of the particle noise on the lensing properties with a 2D convolution. 
For the variance we obtain the following expression,
\begin{equation}
 \sigma^2_Y\xx= \int\!\mathrm{d}^2 \bm x' \tilde\sigma_\Sigma^2(\bm x') \, W_{\sigma^2_Y}(|\bm x - \bm x'|,l_{\mathrm{eff}}) 	
\label{eq:variance_Y}
\end{equation}
and for a single particle noise realisation,
\begin{equation}
 \Delta Y\xx= \int\!\mathrm{d}^2 \bm x' R(\bm x',\tilde\sigma_\Sigma) \, W_Y(|\bm x - \bm x'|,l_{\mathrm{eff}}),
\label{eq:delta_Y}
\end{equation}
where $Y \in \left\lbrace \kappa,\bm\alpha,\bm \gamma \right\rbrace$, $ W_Y(|\bm x - \bm x'|,l_{\mathrm{eff}})$ is the appropriate kernel
and $R(\bm x',\tilde\sigma_\Sigma)$ is a random number drawn from a Gaussian distribution with standard deviation $\tilde\sigma_\Sigma$.
Eq.~\eqref{eq:delta_Y} can be understood as placing uncorrelated particles of size $l_{\mathrm{eff}}$ with random mass $R(\bm x',\tilde\sigma_\Sigma)$ at the 2D positions $\bm x'$.

\subsection{Scaling with \texorpdfstring{$\Npart$}{Npart} and \texorpdfstring{$\Nngb$}{Nngb}} 
\label{subsec:scaling2}
With this effective 2D formulation of the adaptive 3D smoothing we can study the scaling of the noise.
An increased resolution of a N-body simulation samples the density with more particles of smaller masses.
This increase in particle number also decreases the particle noise due to the finer discrete sampling.
With an increased particle number, the number density of particles in 3D is also increased.
Therefore the size of the smoothing kernel, which here is the distance to the $\Nngb$-th neighbour, also changes. 
From Eq.~\eqref{eq:approx_sigmas_sigma} we calculate the scaling of the noise on the surface mass density between different resolutions $k$ and $j$ of a N-body simulation as
\begin{equation}
 \left(\frac{\sigma^{j}_\Sigma\xx}{\sigma^{k}_\Sigma\xx}\right)^2 = \frac{m^j}{m^k} \left(\frac{l^k\xx}{l^j\xx}\right)^2.
\label{eq:scaling_relations}
\end{equation}
If all particles have the same mass, then the fractional change in particle mass is equal to the inverse change in particle number, 
$m^j/m^k = \Npart^k/\Npart^j$. 
The change in the smoothing of the particles, $l$, can be estimated from the change in particle number and the change in the number of neighbours by
\begin{equation}
 \frac{l^k\xx}{l^j\xx} = \frac{l^k}{l^j} \approx \sqrt[3]{\frac{\Nngb^k \Npart^j}{\Nngb^j\Npart^k}}
\label{eq:fraction_smoothing}
\end{equation}
With these two scalings, the fractional change in the noise of the surface mass density in Eq.~\eqref{eq:scaling_relations} simplifies to
\begin{equation}
\left(\frac{\sigma^{j}_\Sigma}{\sigma^{k}_\Sigma}\right)^2
 = 
\left(\frac{\Npart^k}{\Npart^j}\right)^{1/3	}
\left(\frac{\Nngb^k}{\Nngb^j}\right)^{2/3} 
\label{eq:scaling_relations2}
\end{equation}
This first result allows us to estimate the change in the variance of the surface mass density with changing particle number of the simulation 
and a changed smoothing length to convert the N-body particles into lensing properties.

We can transform Eq.~\eqref{eq:approx_sigmas_sigma} to
\begin{equation}
    \frac{\sigma^2_\Sigma\xx}{\Sigma^2\xx} \approx \frac{9}{5\pi}\frac{m_p}{l^2\xx A} \frac{1}{\Sigma\xx}.
  \label{eq:sigma_mass_paper}
\end{equation}
For the special case of a uniform density field in 3D with equally distributed particles, 
the mean projected 2D surface mass density is $\bar\Sigma = m_p \Npart A / L^2$, 
where $L$ is the side length of the cube of equally distributed particles.
For this configuration, the distance $l$ to the $\Nngb$-th neighbour can be calculated from 
\begin{equation}
\frac{l}{L} = \left(\frac{\Nngb}{\Npart} \right)^{1/3}.
\end{equation}
Substituting these two relations into Eq.~\eqref{eq:sigma_mass_paper} we obtain in units of the critical density $\Sigma_{\mathrm{crit}}$, the following relation

\begin{equation}
 \frac{\sigma_\kappa\xx}{\bar\kappa\xx} = \sqrt{\frac{9}{5}} \frac{1}{\Nngb^{1/3} \Npart^{1/6}}
\end{equation}
which is similar to what (\cite{2006ApJsmoothing}) found by numerical fitting (their Eq.~(4)). 
The exact value of the  proportionality constant depends on the exact form of the kernel.

\section{Comparison of the Particle Noise with Substructure}
\label{sec:Comparison_of_the_Particle_Noise_with_Substructure} 

In this second part of the paper we compare the small-scale fluctuations due to two competing effects, 
the particle noise and physical mass substructures (subhalos). 
In particular, we are interested in quantifying the limit at which the effect of physical mass substructure becomes comparable to that of particle noise, 
and the substructure is considered too `small' to be `visible' above the noise level. 
To this end, we need quantitative measures of the effect of a substructure on the lensing properties. 
Mainly, we need a metric to answer the following questions: which lensing property is best to look at in order to compare the substructure to the noise?
Which property of the substructure is the best one to quantify how `small' a substructure is? 
When can a substructure be considered `visible' above the noise level?

Substructures in our N-body simulation are identified as gravitationally bound objects consisting of more than $N_{\mathrm{min}} \sim 20$ particles. 
For each substructure we can measure, among other parameters, its mass, size, ellipticity, density profile and circular velocity.
For a simplified substructure model such as a singular isothermal sphere (SIS), the lens strength $b$ is proportional to the Einstein radius 
which is proportional to the one-dimensional velocity dispersion $\sigma_v^2$. 
The mass within $R$ on the other hand is proportional to $\sigma_v^2 R$. 
Therefore we choose the quotient of the mass of the substructure and its half-mass radius in units of the critical density, 
$\beta = M_{\mathrm{ss}} / R_{\mathrm{HMR}} / \Sigma_{\mathrm{crit}}$ as a measure of the strength of each substructure. 
We will confirm in Sec.~\ref{sec:lens_strength} that this measure derived from a simplified SIS model is a good parameter to quantify the effect of the numerically simulated substructures.

For the convenience of the reader we will also introduce a second x-axis on the top of Figs.~\ref{fig:detection_alpha} to \ref{fig:required_resolution}.
This second axis converts the lens strength on the bottom x-axis to a typical substructure mass, $M_{\mathrm{ss}}$.
The conversion is a linear fit to all numerically simulated substructures as identified by SUBFIND within the Level-2 simulation of cluster E
in $\log(M_{\mathrm{ss}})$-$\log(\beta)$ space within $-2<\log_{10}\beta_{\mathrm{ss}}<0$. This corresponds
roughly to $0.03 < M_{\mathrm{ss}}/(10^{10}\Msun) < 7 $.
The linear fit, $\log(M_{\mathrm{ss}}/(10^{10}\Msun/h)) = 1.21^{\pm 0.01} \log_{10}(\beta_{\mathrm{ss}}/\arcsec) + 0.71^{\pm 0.02}$,
averages different substructure profiles, sizes and concentrations and therefore yields a typical substructure mass for a given substructure lens strength. 
All calculations, however, are performed in terms of the substructure lens strength $\beta$ in order to fully take into account the different 2D profiles of the simulated
substructures.

\subsection{Substructure Surface Mass Density}
\label{subsec:Substructure_Surface_Mass_Density}
As a first, although very simple step, one might compare the projected scaled surface mass density of each 
substructure, $\kappa_{i}$, to the amplitude of the noise fluctuations of the surface mass density due to the particle noise, $\sigma_{\kappa}$.
\begin{figure}
\begin{center}
\includegraphics[width=1.0\columnwidth]{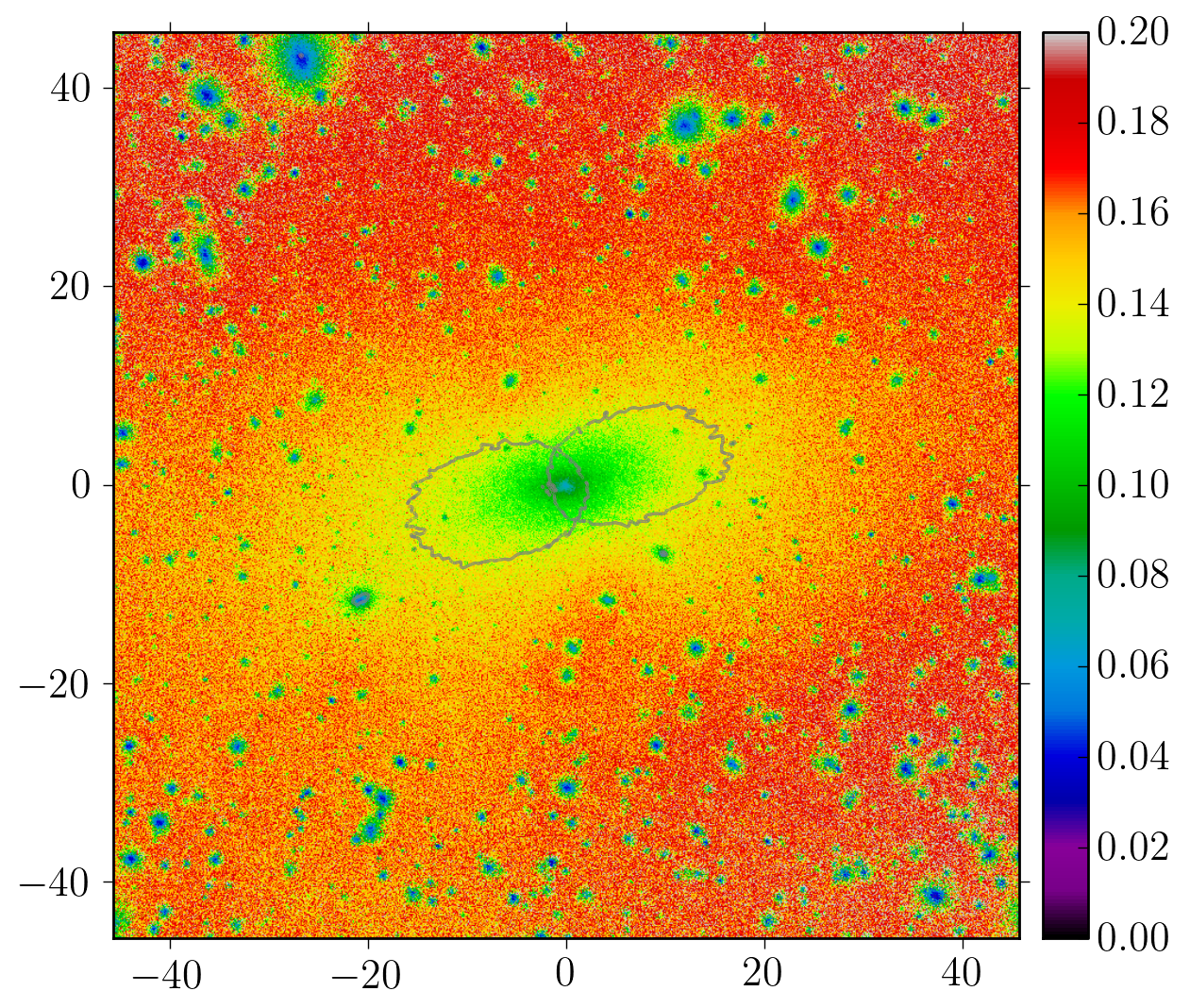}
\end{center}
\caption{Approximation of the lens strength $\beta$ of Gaussian noise with the same parameters as the noise on the surface mass density for the Level-2 version of cluster E.
At the critical lines the lens strength of the $3\sigma$ noise is $\beta\sim 0.14$. 
}
\label{fig:noise_lens_strength}
\end{figure}
We approximate the noise by a Gaussian fluctuation field with amplitude $\sigma_\kappa$ 
and a correlation length equivalent to the 2D smoothing length, $\leff$, at each point on the lens plane. 
We can then calculate the equivalent lens strength for the noise fluctuations.
The size of the Gaussian is $\leff = l$, the total mass of a $n\sigma$ noise substructure
is therefore $\pi l^2 n\sigma_\Sigma$ for $n \in \{ 1,2,3\}$. 
The half mass radius is $R_{\mathrm{HMR}} = l \sqrt{2\ln{2}}$.
Therefore the equivalent lens strength is $\beta = M_{\mathrm{ss}}/(R_{\mathrm{HMR}} \Sigma_{\mathrm{crit}}) = \pi l \sigma_\kappa / (2\ln{2})$.
Figure \ref{fig:noise_lens_strength} shows the lens strength of the noise on the scaled surface mass density at each point on the lens plane. 
At the critical line, we therefore identify the substructures that will have a smaller influence than the $3\sigma$ noise as 
those substructures which are smaller than $\beta \sim 0.14$, this corresponds to an average subhalo mass smaller than $\Mss \sim 6.5 \ten{9} \Msun $.

By comparing the surface mass density, we are essentially comparing a single substructure with 
a random field of positive and negative substructures for the noise. 
Since the lensing equation is highly nonlinear, especially in the strong lensing regime, 
the noise also propagates nonlinearly through the lensing properties as we will show in the following sections. 
We can therefore not assume that the limits derived here based on the surface mass density are the true resolution limits of the simulation,
in the sense that every lensing property yields identical limits.

\subsection{Substructure Magnification}
\label{subsec:Substructure_Magnification}
\begin{figure}
\begin{center}
\includegraphics[width=1.0\columnwidth]{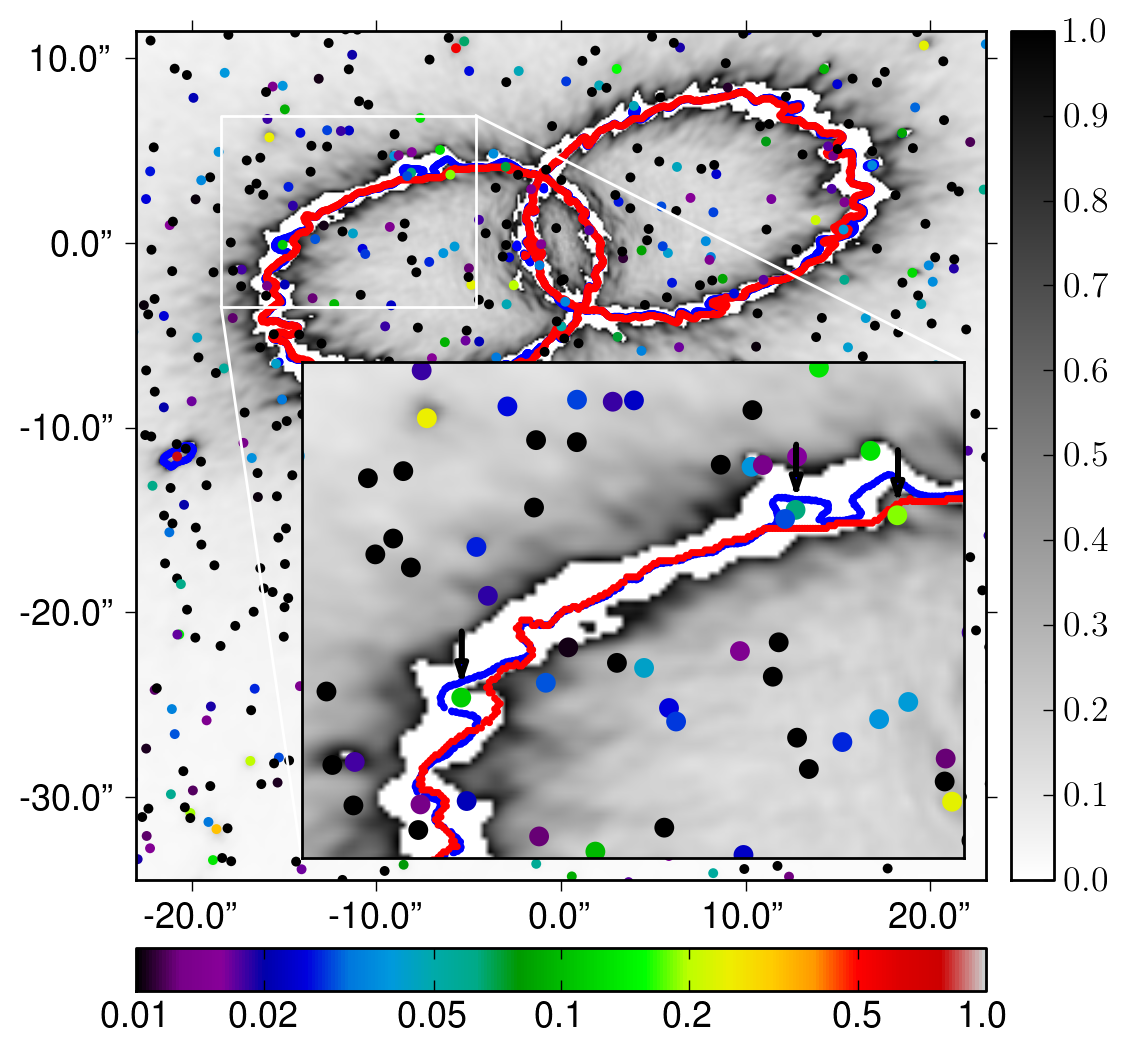}
\includegraphics[width=1.0\columnwidth]{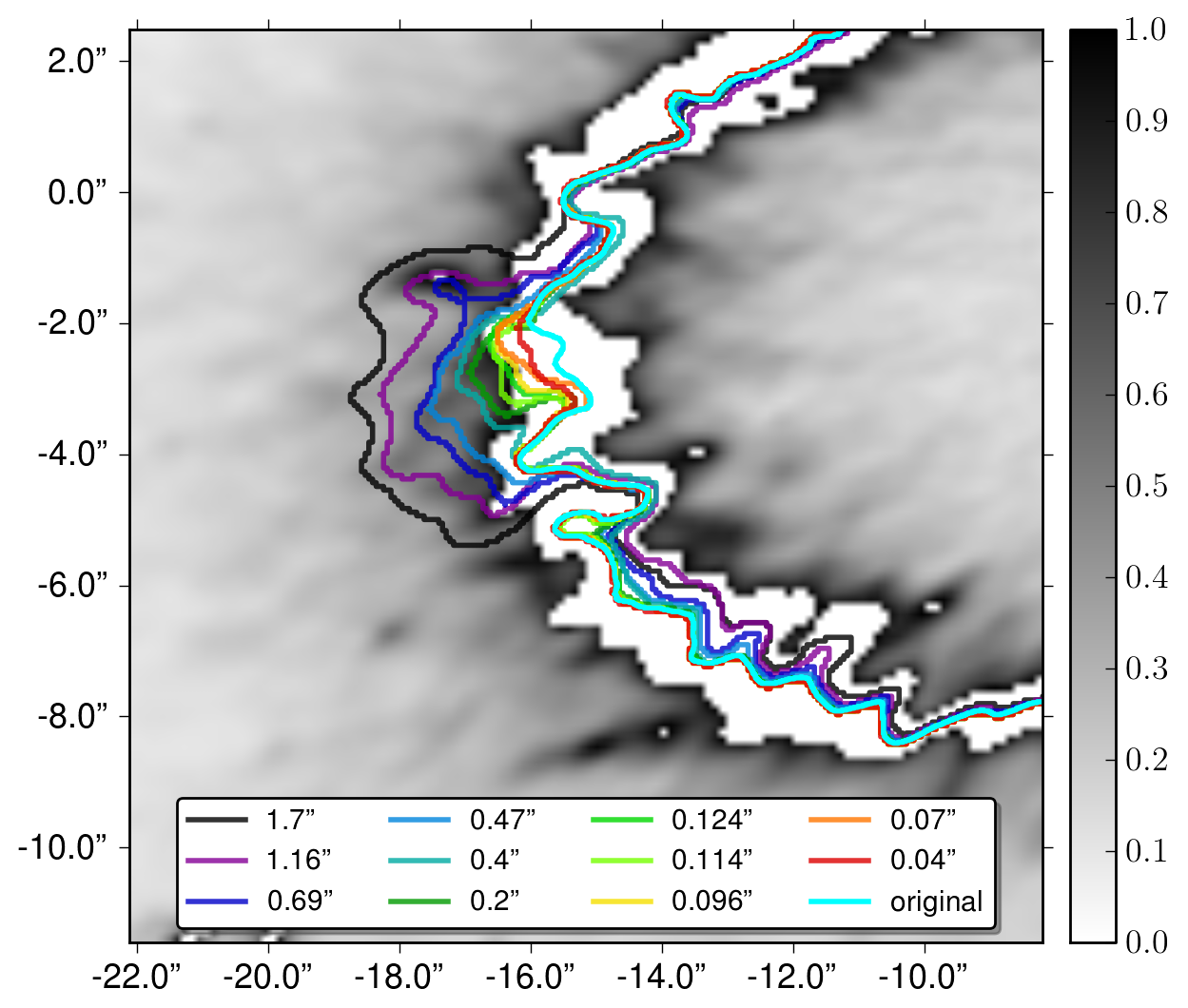}
\end{center}
\caption{Top panel: Critical lines with (blue) and without (red) subhalos. 
Background in grey scale is the particle noise of the inverse of the magnification $3\sigma_{\mu^{-1}}/|\mu^{-1}|$. 
The noise on the inverse of the magnification is cut-off at $3\sigma_{\mu^{-1}} \sim |\mu^{-1}|$ (white band following the critical curves),
this indicates the noise on the critical line. Coloured circles indicate substructures with masses in $10^{10} \Msun$.
The inset shows an enlargement with the significant effect of three (four) subhalos directly on the critical line marked by arrows, for details see text.
Bottom panel: 11 substructures of decreasing lens strength artificially added on top of the N-body cluster mass distribution at $-15.6 \arcsec, -2.6 \arcsec$.
Substructures with a lens strength smaller than $\sim 0.1\arcsec$ are within the $3\sigma$ limits of the noise on $\mu^{-1}$ shown as a grey scale 
background.
}
\label{fig:noise_magnification_ss}
\end{figure}
As a second quantity we compare the effect of mass substructure and particle noise on the inverse of the magnification $\mu^{-1}$ and on the critical lines. 
Both panels in Figure \ref{fig:noise_magnification_ss} show the particle noise on the inverse of the magnification $3\sigma_{\mu^{-1}}/|\mu^{-1}|$ as a grey scale background. 
The noise increases with the magnification and reaches its maximum at the critical lines. In order to indicate the width of the $3\sigma$ `wiggles' in the critical line, 
values where the noise $3\sigma_{\mu^{-1}}$ exceeds $|\mu^{-1}|$ close to the critical curves are set to white.
In the top panel, two critical lines, $\mu^{-1} \rightarrow 0$, are over plotted. 
The blue line is for the original N-body cluster E from the Level-2
Phoenix simulations. The red line is for the same cluster but with all subhalos identified by SUBFIND removed. 
The physical subhalos of the simulation are marked as coloured circles. The colour indicates the subhalo lens strength $\beta = M_{\mathrm{ss}} / R_{\mathrm{HMR}} / \Sigma_{\mathrm{crit}}$.
The two critical lines are almost identical except for those few cases in which a subhalo lies directly on top of the critical line. 
The inset is showing an enlargement of three of these cases indicated by arrows.
The masses of the three subhalos are 6.0 (green on the left), 2.9+0.7 (top middle, green+blue) and 8.4 (top right, light green) $\times 10^9\Msun$. 
The influence of the two more massive subhalos in the inset with masses of 6.0 and 8.4 $\times 10^9 \Msun$, exceeds the $3\sigma$ noise of the critical line.
It is evident from the critical line of the N-body cluster without any subhalos, that the curve still shows a lot of irregularities. 
Those wiggles are all a consequence of the particle noise due to the discrete N-body representation with finite size particles.

In order to study the influence of substructures of different sizes we could rotate the cluster, 
however, this  would also change the overall shape of the critical lines and make the comparison between different substructure difficult to quantify. 
A better approach is then to artificially place additional substructures on top of the critical lines.
In the bottom panel of Fig.~\ref{fig:noise_magnification_ss} we quantify the effect of artificial substructures. 
We add the substructure particles of 11 substructures randomly chosen with decreasing lens strength on top of the particle distribution 
of the original N-body simulation at $(-15.6 \arcsec, -2.6 \arcsec)$.
For each substructure we calculate the new critical lines and compare the deviation from the original critical line with the $3\sigma$ noise on the 
inverse of the magnification $\mu^{-1}$ in the lower panel of Fig.~\ref{fig:noise_magnification_ss}. 
Substructures smaller than $\beta \approx 0.1\arcsec$ are within the white band in the lower panel of Fig.~\ref{fig:noise_magnification_ss}.
This limit corresponds to resolved minimum average subhalo mass of $\Mss \sim 4.3 \ten{9} \Msun$.
Any substructure bigger than this will cause a `wiggle' in the critical line that is stronger than those caused by the numerical $3\sigma$ particle 
noise of the simulation. 

\subsection{Substructure Deflection Angles}
\label{subsec:Substructure_Deflection_Angles}
In this section, we compare the deflection angles of the substructures $\bm \alpha_{\mathrm{ss}}$ with the noise on the deflection angles due to the particle noise in the 
N-body simulation, $\sigma_{\bm \alpha}$. Although the deflection angles are not directly observable, we will show in this section that they are a good measure of the effect of a substructure.

We have already seen in Sec.~\ref{sec:Images}, that the additional deflection caused by 
small-scale fluctuations due to the particle noise can be approximated as a small correction on top of the deflection by the numerical cluster. 
Any additional deflection $\Delta \bm \alpha$ will result in a change in the observable surface brightness of the image $\Delta \bm d$. 
In a simplified model where the source gradient is varying slowly, 
the greater the additional deflection $\Delta \bm \alpha$, the greater the change in the image brightness at that point (see Eq.~\eqref{eq:lin_delta_d}).
Therefore we compare the additional deflection by a substructure, $\Delta \bm\alpha_{\mathrm{ss}}$, with the fluctuations in the deflection angles, 
$\sigma_{\bm \alpha}$ caused by the particle noise.
We could also use an integrated measure over all points of the image plane, where the additional deflection by the substructure exceeds 
the magnitude of the fluctuations from the particle noise. 
Since the results are the same, we use here for simplicity only the maximum value of 
the additional substructure deflection, $\mathrm{max}[\Delta \bm\alpha_{\mathrm{ss}}\xx]_{\bm x}$
and compare it to $\sigma_{\bm \alpha}\xx$ due to the particle noise at the same point on the lens plane.	

\begin{figure}
\begin{center}
\includegraphics[width=1.0\columnwidth]{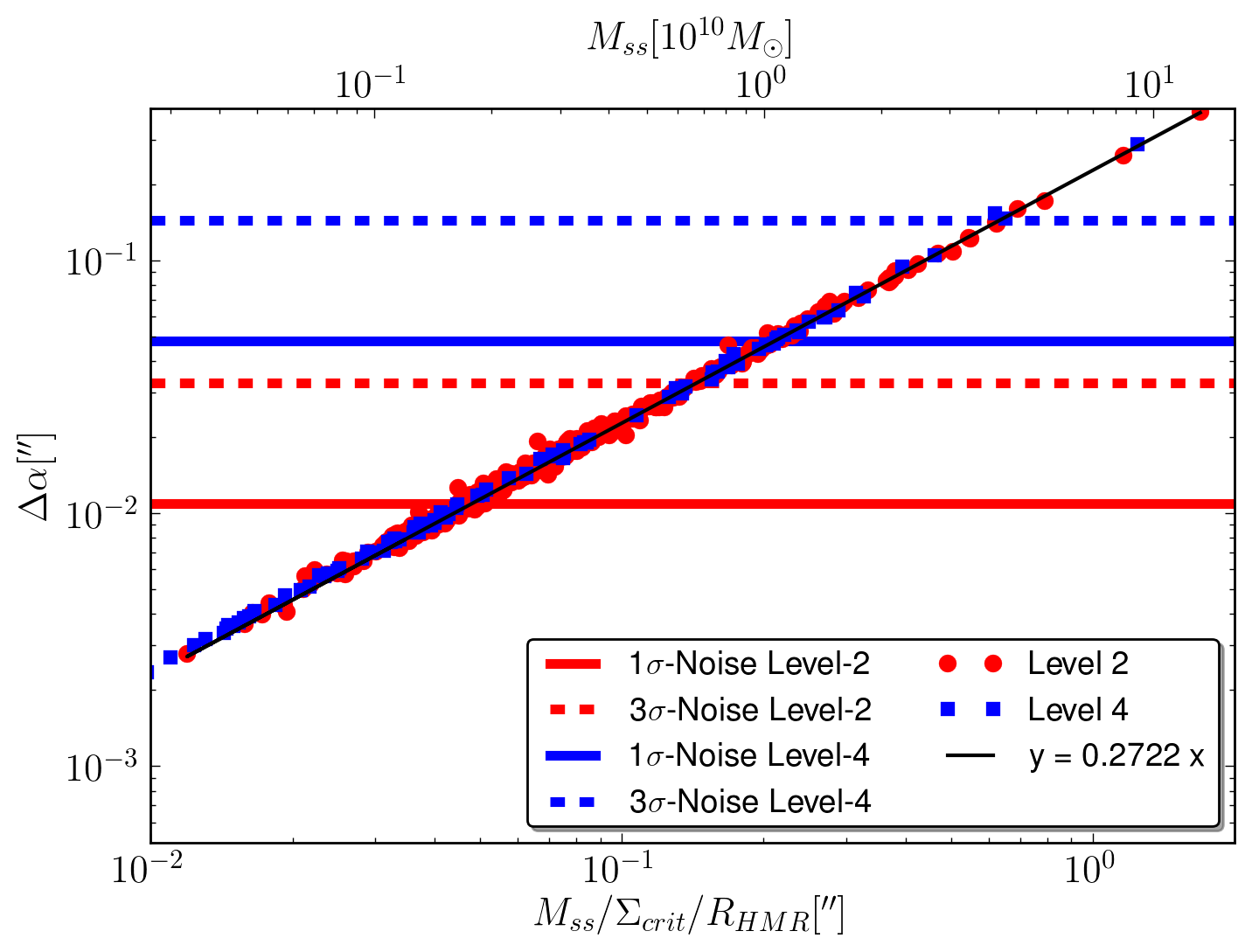}
\end{center}
\caption{ Maximum value of the additional deflection $\Delta \bm \alpha$ for 
substructures with positions within $(92\arcsec)^2$ from the centre for the populations
Level-2 and Level-4 resolutions. Horizontal lines show the $1\sigma$ (solid lines) and the $3\sigma$ (dashed lines) noise at the position of the lowest image in the top left panel of 
Fig.~\ref{fig:images_constant_source} for Level-2 (red) and Level-4 (blue).
}
\label{fig:detection_alpha}
\end{figure}
Figure~\ref{fig:detection_alpha} shows the maximum of the additional deflection for a subsample of the 2597 
substructures in the central $(92\arcsec)^2$. The subsample is chosen to include the most massive substructures with the 
greatest lens strength. 
Each substructure is shown as a red circle for the Level-2 resolution.
As a comparison, the subhalos of a second, lower resolution of the simulation, Level-4, are also shown as blue squares.
We use the fully numerical substructures to calculate the total mass and half mass radius for the lens strength on the x-axis.
The y-axis is the maximum of the additional deflection calculated numerically by solving the Poisson equation for each substructure. 
All subhalos fall almost perfectly onto the linear relation $y = 0.2272 x$, \label{sec:lens_strength}
therefore the parameter $M_{\mathrm{ss}}/(\Sigma_{\mathrm{crit}} R_{\mathrm{HMR}})$ is suitable to 
quantify the strength of the numerical substructures and 
the subsample of 400 subhalos is sufficient to quantify the the effects of the subhalo population in Fig.~\ref{fig:detection_alpha}.
Now, we compare the maximum additional deflection caused by the individual substructures with the noise on the deflection angles 
from Sec.~\ref{sec:Deflection_Angles} as follows.
We artificially place each of the substructures directly behind the lowest image at $(13\arcsec, -4\arcsec)$.
Since the noise on $\bm \alpha$ is a very smooth function over the size of a typical image (see Fig.~\ref{fig:noise_alpha}) 
we can use the same value $\sigma_{\bm \alpha}$ for all substructures.
The noise on the deflection angles can therefore be represented by horizontal lines in Figure~\ref{fig:detection_alpha}. 
A more rigorous approach would require to use the position of the respective maximum of the additional deflection for each substructure 
which varies slightly with the size of the substructure. 
Using the correct appropriate values of $\sigma_{\bm \alpha}$, however, does not alter the results. 
Under the assumption that a greater additional deflection by a substructure, $\Delta \bm\alpha_{\mathrm{ss}}$, also causes a more significant change in the image brightness, 
we define as visible substructure, those that are above the respective horizontal noise levels in Figure~\ref{fig:detection_alpha}.

With this method we are able to constrain the $1 \sigma$ `visibility' of substructures to a substructure lens strengths
of $0.048\arcsec$ $(1.7\ten{9}\Msun)$ for Level-2 and $0.21\arcsec$ $(1\ten{10}\Msun)$ for Level-4 (solid lines).
The respective $3\sigma$ limits are $0.14 \arcsec$ $(6.5\ten{9}\Msun)$ and $0.62 \arcsec$ $(3.9\ten{9}\Msun)$ (dashed lines). 
This simple comparison of the deflections by a substructure with the fluctuations in the deflection from the particle noise 
provides a fast measure of the detectability of a substructure. 
We will see in the next section that these results based on the deflection angles are almost identical 
to the analysis based on differences in the image brightness distribution.

\subsection{Lensed images with Substructure of a Gaussian Source}
\label{subsec:Substructure_Images}
\begin{figure}
\begin{center}
\includegraphics[width=1.0\columnwidth]{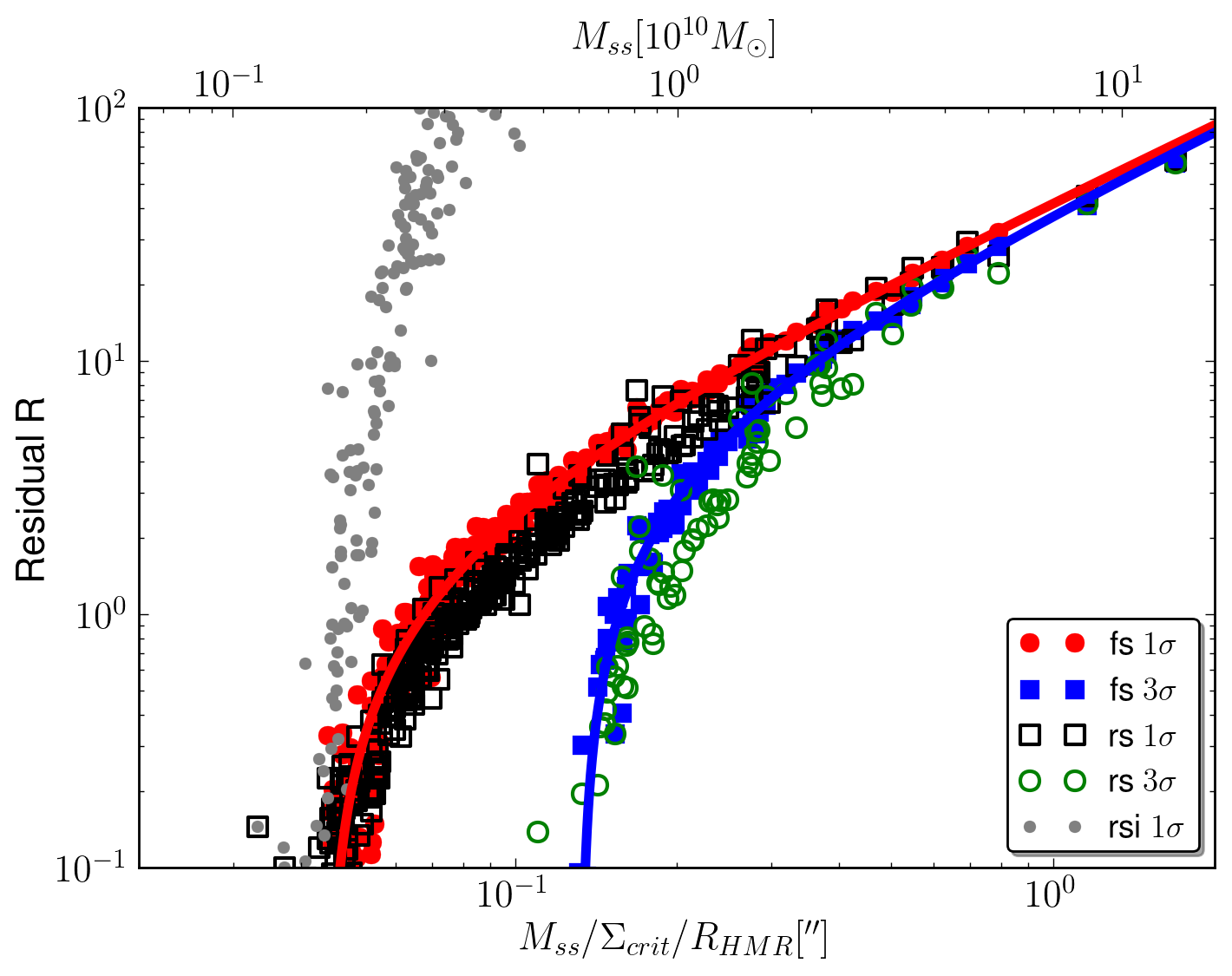}
\end{center}
\caption{ Residual $R$ where $R = |\Delta \bm d| - n\sigma_{\bm{d}} > 0$, where the image brightness difference caused by adding a substructure is bigger than 
the particle noise induced brightness difference in the image. 
Shown is the maximum of R measured over all points on the lens plane. Points are for $>2500$ substructures from the central $(92\arcsec)^2$ of the cluster artificially
placed on top of the lowest image from the top left panel of Fig.~\ref{fig:images_constant_source}.
Filled symbols are for a fixed source brightness distribution (fs) red circles for $1\sigma$ and blue squares for $3\sigma$ noise, 
solid lines are a linear fit to the $1$ and $3 \sigma$ points.
Open symbols are for a reconstructed source (rs) where the only assumption is the regularity of the source, black squares for $1\sigma$ and green circles for $3\sigma$.
Grey small filled circles are the integrated residual over the image plane with a reconstructed source (rsi) and $1 \sigma$ noise.
The lower limits of resolved substructure lens strengths derived from the linear fits are $0.045\arcsec (1\sigma)$ and $0.13\arcsec (3\sigma)$.
}
\label{fig:detection_images}
\end{figure}
In this section we compare the effect of substructures within the N-body simulation with the particle noise based on the surface brightness distribution of the images. 
The gravitationally lensed images are the only directly observable quantities, 
therefore any observational detection of a substructure in the lens will be based on a reconstruction of the observed image positions or the image brightness distribution.
If there is no small-scale structure in the lens, it is in theory possible to find a source surface brightness distribution that fits all of the multiply lensed images.
The presence of a substructure in the lens close to one image of a multiply imaged system will result in a local imprint of the substructure on the closest image. 
Therefore, no source brightness distribution exists, that, when lensed through a smooth large-scale lensing mass distribution, will be able to model all multiply lensed images simultaneously. 
Therefore, there has to be substructure in the lens.

Here, we are using a similar method to quantify the effect of a substructure.  
We artificially place a substructure on the lens and simulate an image with substructure.
We then try and fail to reconstruct this lens with substructure with a smooth model. 
In our case, we know the underlying smooth model, which is the same lens without the substructure.
We can therefore calculate the image for the idealised smooth model by lensing a source through the lens model without the added substructure. 
The difference between the image for a lens with substructure and for a lens without substructure 
is then a measure for the failure of the smooth model to reproduce the image with substructure.

Similarly to Sec.~\ref{sec:Images}, we use two different approaches to evaluate the influence of substructure on an image brightness distribution. 
The first method uses a fixed source surface brightness distribution to create a reference image $\bm d_0$ using the N-body cluster lens without any substructures close to any of the lensed images.
This reference image is shown in the top left panel of Fig.~\ref{fig:images_constant_source}.
We then use the 2597 substructures from the central $(92\arcsec)^2$ as a sample of numerically simulated substructures and artificially place each substructure on top of the 
cluster directly behind the lowest of the three images at $(13\arcsec, -4\arcsec)$.
At this position on the lens plane the substructure will have the biggest effect. 
For each substructure we then lens the same source and obtain 2597 different images $\bm d_i$. 
For each of these images we calculate the image brightness difference $\bm d_i - \bm d_0 = \Delta \bm d_i$ 
at each point on the lens plane. This is the brightness difference due to the artificially added substructure. 
We then compare this brightness difference $\Delta \bm d_i$ with the discreteness noise on the image $\sigma_{\bm{d}}$ from Sec.~\ref{sec:Images}.
We calculate the residual, $R_i$, as a difference in the image brightness due to the substructure that is greater than 
the amplitude of the fluctuations in the image surface brightness distribution at that point due to the particle noise as
\begin{equation}
  R_i\xx = |\Delta \bm d_i\xx| - n \sigma_{\bm{d}}\xx \geqq 0 \qquad n = \left\lbrace1,2,3\right\rbrace. 
\label{eq:residual}
\end{equation}
The results for a residual integrated over the lens plane are identical to the results from the comparison based on the maximum value $\mathrm{max}[R_{i}\xx]_{\bm x}$ shown as 
an example for the integrated $1\sigma$ residual (rsi) multiplied by a factor of $0.01$ to enhance the contrast in Fig.~\ref{fig:detection_images}.
The maximum value of the residual, $R_{i}$,  is shown as solid red circles $(1\sigma)$ and solid blue squares $(3\sigma)$ 
in Fig.~\ref{fig:detection_images} for the Level-2 resolution using a fixed source surface brightness distribution (fs). 
The y-axis is the maximum residual $R_{i}$ and the x-axis is the lensing strength of the substructure.
The solid lines are linear fits that constrain the visible substructures. 
We consider a substructure to be visible when $R_i > 0$. 
The limits are  $0.045 \arcsec$ $(1\sigma)$, $0.09 \arcsec$ $(2\sigma)$ and $0.133 \arcsec$ $(3\sigma)$
corresponding to 1.6, 3.8 and 6 $\ten{9}\Msun$ respectively for a fixed source brightness distribution.
Any substructure with a lens strength greater than these lower limits placed on top of the lowest image will result in an image brightness difference
that is greater than the fluctuations due to the particle noise.
These limits are very close to the values derived from the deflection angles in the previous section.

We have seen in Sec.~\ref{sec:Images} that lensing one fixed source surface brightness distribution through different noise realisations of the lens 
results in an overestimation of the noise on the image by a factor of $\sim 2$. This definition of the noise on the image 
also includes an artificial relative source-caustic shift (see Sec.~\ref{sec:Images}  for details), we 
therefore  expect that the assumption of a fixed source here also is an over simplification. 
We therefore use the method described in Sec.~\ref{sec:Images} to reconstruct the 
best possible source (and therefore also the closest image) using the deflection angles with substructure to match the same image without substructure. 
For each of the reconstructed substructure images we then calculate the residual Eq.~\eqref{eq:residual} from the previous paragraph. 
The maximum of these residuals are shown in Fig.~\ref{fig:detection_images} with black open squares $(1\sigma)$ and green open circles ($3\sigma$) for a reconstructed source (rs),
as well as an integrated residual as an example (rsi $1\sigma$).
The points no longer follow a linear relation.
But the lower limits derived with this method 
of $0.045 \arcsec (1\sigma)$ and $0.13 \arcsec (3\sigma)$ if we reconstruct the source for each substructure image
are identical to the ones with a fixed source brightness distribution.

This nontrivial result shows that including the unobservable relative source-caustic shift in the noise increases the noise by about a factor of two, see 
Sec.~\ref{sec:Images}. But at the same time it also increases the image brightness difference due to a substructure in the lens.
Therefore the cutoff $R=0$ which indicates the minimum size of the resolved substructures remains unchanged.
This is very convenient, since it allows us to accurately calculate the limits with much simpler and faster methods without having to reconstruct the 
source for each of the substructures and bootstrapped particle noise realisations of the cluster.
The limits for the smallest resolved substructures that we found in this section based on the image brightness distribution of the lensed images
are very similar to the limits from simpler lensing properties such as the deflection angles in Sec.~\ref{subsec:Substructure_Deflection_Angles} or the 
surface mass density in Sec.~\ref{subsec:Substructure_Surface_Mass_Density}.

Up to here we have used a Gaussian source surface brightness distribution. 
This simplified source allows for a systematic description of the the effect of the particle noise on the lensed images, 
however we expect real source galaxies, especially at $z=2$, to be more structured. In the following, therefore, we simulate sources with different degrees of structure.

\subsection{Lensed images with Substructure of a Realistic Source}
\label{subsec:Substructure_Images_Realistic_Source}
In order to simulate various degrees of smoothness for the source surface brightness distribution, 
we use the very structured source surface brightness distribution of a true galaxy as observed with HST.
We then smooth this source surface brightness distribution with a Gaussian kernel with increasing sizes denoted as 20, 40 and 100. 
As an example, the source brightness distributions for 100 and 20 are shown in the bottom panels of Fig.~\ref{fig:images_noise_different_sources}.
For each different source we recalculate the noise on the image brightness distribution, $\sigma_d$, as described in Sec.~\ref{sec:Images}.
\begin{figure}
\begin{center}
\includegraphics[width=1.0\columnwidth]{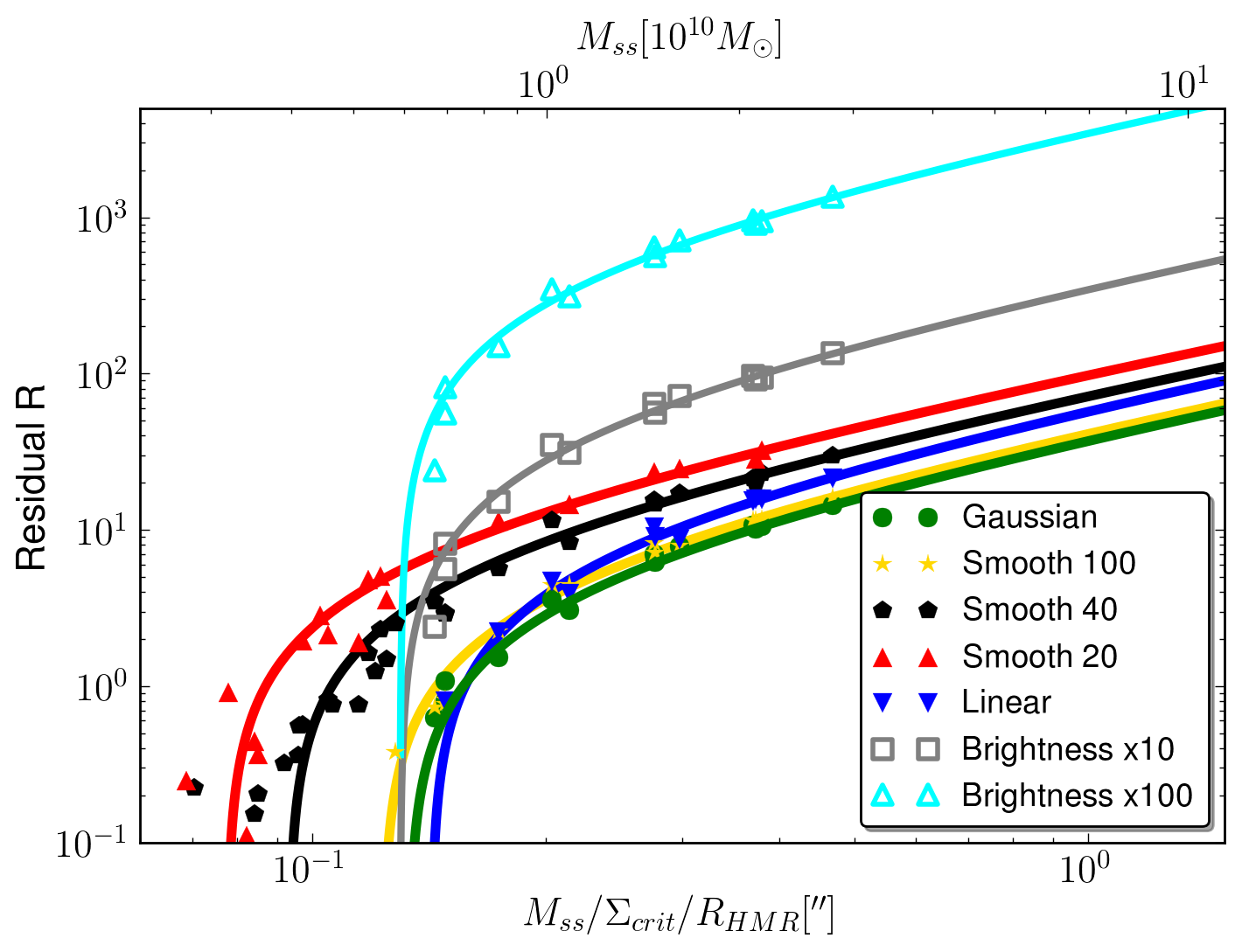}
\end{center}
\caption{ The same as in Fig.~\ref{fig:detection_images} but for different source brightness distributions and $3\sigma$ residuals. 
A very structured source brightness distribution
is smoothed with a Gaussian kernel of sizes 20, 40 and 100, as a comparison, the Gaussian source surface brightness distribution from Fig.~\ref{fig:detection_images} and 
a linear source surface brightness distribution are also shown.
A more structured source shifts the cutoff towards smaller substructures, while the overall
residual is increased. Increasing the brightness of the source by a factor of 10 and 100 
does not change the smallest resolved substructures, but increases the residual.
For the sake of the clarity of the figure, only every 5th substructure is plotted for each line. 
}
\label{fig:detection_images_brightness_structure}
\end{figure}
We show the residual above the particle noise on the image brightness, of Eq.~\eqref{eq:residual} for $n=3$, in Fig.~\ref{fig:detection_images_brightness_structure}.
As the structure in the source is decreased, an increasingly smaller number of substructures can be resolved above the noise limit, and 
the cutoff where $R \rightarrow 0$ shifts to the right. The limit for resolved substructures for the smoothest source, 100, 
is almost identical to the limit calculated with a Gaussian source surface brightness distribution in the previous section. 
In Fig.~\ref{fig:detection_images_brightness_structure} we additionally show the residual using two different sources with an
increased maximum source surface brightness by factors of 10 and 100 and the original Gaussian brightness distribution. 
As an extreme and unphysical limit we also show the residual for a source surface brightness distribution that is decreasing linearly with increasing
distance from the source centre.
Size and position of the source are kept constant for each of the seven curves. The behaviour of the curves can be qualitatively understood as follows.

Changing the brightness of the source while keeping the source size constant, increases the gradient of the source. 
Therefore, the same change in the deflection angles, $\Delta \bm \alpha$, 
due to either noise or substructure, will result in a greater change in image brightness, $\Delta \bm d \approx \Delta \bm \alpha \cdot \nabla s$, 
see also Eq.~\eqref{eq:lin_delta_d}.
This effect, however, affects the calculation of the noise on the images, $3\sigma_{\bm d}$, and the image brightness difference for each substructure, 
$\Delta \bm d_k$, in the same way. Therefore even though the residual Eq.~\eqref{eq:residual} is increased, the lower limit of resolved substructures is unchanged.
To understand the behaviour of the curves with increasing source structure, we consider a small substructure that is barely not resolved 
with a Gaussian source brightness distribution, for example a substructure with a lens strength of $0.12\arcsec$ which corresponds to a substructure mass of $\sim 5.4 \ten{9}\,\Msun$. 
Since the residual of this substructure is $R_{\mathrm{max}}\sim 0$, 
the image brightness difference at any point due to this substructure, $\Delta \bm d_i$, 
is comparable to the noise on the image brightness, $3\sigma_{\bm d}$, which we calculated from the different 
particle noise realisations of the cluster,  $\sigma_{\bm d} = 1/(N-1) \sum_k \Delta \bm d_k$. 
If we now increase and change the form of the source gradient with a more structured source on scales of $\Delta \bm\alpha_i$,
the same change in the deflection angles due to the substructure, $\Delta \bm \alpha_i$, will result in an increased change in the image brightness, $\Delta \bm d_i$.
In contrast to the increased source brightness, which only changes the amplitude of the source gradient, 
a more structured source additionally changes the shape of the gradient on scales smaller than $\Delta\bm\alpha_i$. 
Therefore and because the noise $\sigma_{\bm d}$ is a nonlinear combination of the effects of a field of positive and negative noise substructures 
it behaves differently than a more coherent single substructure.
As we can see from Fig.~\ref{fig:detection_images_brightness_structure} the net effect is more prominent for a single substructure. 
An increased source structure therefore allows the detection of smaller substructures above the particle noise limit.

Fig.~\ref{fig:detection_images_brightness_structure} shows that 
a Gaussian source surface brightness distribution is a bad choice in terms of the lower limit of resolved substructures. 
The results are almost identical to the worse case, a linear source brightness distribution 
The truly worst choice, however, would be a constant source brightness which is completely indifferent to changes in the deflection angles. 
A linear source gradient includes both, an increased gradient and at the same time less structure in the source brightness distribution 
with respect to a Gaussian source surface brightness distribution.
Therefore, the overall residual is slightly higher and the cutoff is shifted to greater substructure lens strengths.
The linear source surface brightness distribution is the extreme limit, 
a constant source gradient in Eq.~\eqref{eq:lin_delta_d}. Therefore, the cutoff at $\beta = 0.14\arcsec$ ($6.5 \ten{9}\Msun$) is identical to the lower limit 
derived from the substructure deflection angles in Sec.~\ref{subsec:Substructure_Deflection_Angles}.

\section{Scaling of the Resolution Limit}
\label{subsec:Substructure_Scaling}
\begin{figure}
\begin{center}
\includegraphics[width=1.0\columnwidth]{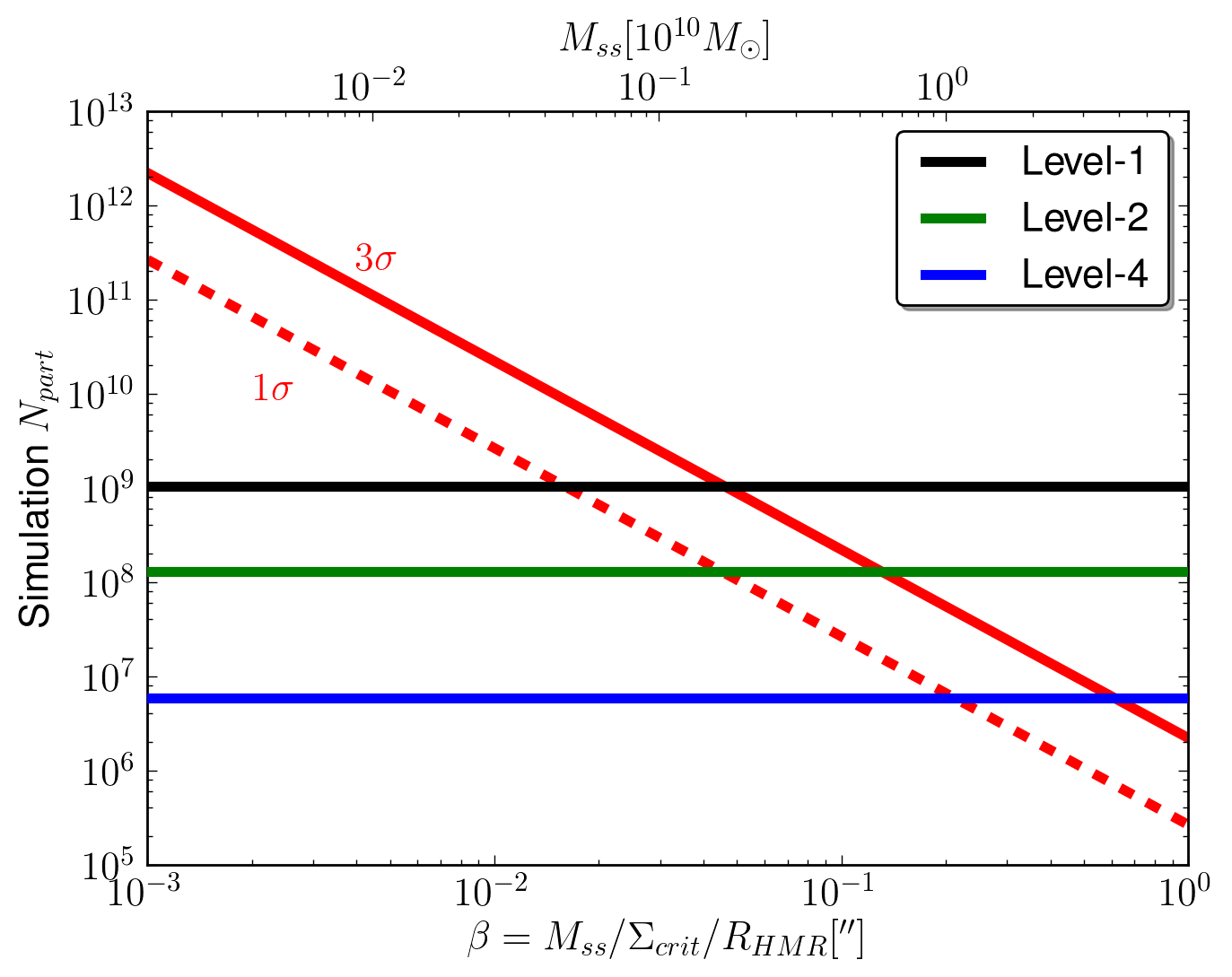}
\end{center}
\caption{ Required number of particles in a N-body simulation targeted to resolve substructures above $3\sigma$ of the noise with a 
given lens strength with strongly gravitational lensed images. The second x-axis converts the lens strength of the substructure 
to the average mass of a simulated substructure.
The horizontal lines show three resolutions of the Phoenix simulations and their respective lower $3\sigma$ limits
of  $\beta_{\mathrm{ss}}^{\mathrm{min}} \sim 0.046\arcsec$ or $M_{\mathrm{ss}}^{\mathrm{min}} \sim 1.7\ten{9} \Msun$ for Level-1
and $\beta_{\mathrm{ss}}^{\mathrm{min}} \sim 0.135\arcsec$ or $M_{\mathrm{ss}}^{\mathrm{min}} \sim 5.8\ten{9} \Msun$ for Level-2 
and $\beta_{\mathrm{ss}}^{\mathrm{min}} \sim 0.6\arcsec$ or $M_{\mathrm{ss}}^{\mathrm{min}} \sim 3.8\ten{10} \Msun$ for Level-4.
}
\label{fig:required_resolution}
\end{figure}
In the previous sections we derived different methods to compare the effect of a substructure on different lensing properties
with the effect caused by the discrete representation of the N-body cluster with particles and we
found lower limits on the substructure lens strength. Substructures with a weaker lens strength do not affect the lensing properties
strong enough in order to be detectable above the particle noise limit.

This allows us to answer a very interesting question.
If we plan to simulate gravitational lensing with a numerically simulated N-body mass distribution, what resolution
do we need, in order to `resolve' substructures of a given size.

The most advanced lower limits derived from the image brightness
distribution with a reconstructing source method are very close to the simplified limits predicted from deflection angle differences, 
or even the simple comparison based on the surface mass density fluctuations in Sec.~\ref{subsec:Substructure_Surface_Mass_Density}. 
We already derived scaling relations for the noise on the surface mass density in Sec.~\ref{sec:scaling}, 
therefore, we can now combine those results. 
If we approximate the noise as positive and negative Gaussian fluctuations, we can write the lens strength of one 
noise fluctuation as $\beta = \pi l \sigma_\kappa / (2\ln{2})$, see Sec.~\ref{subsec:Substructure_Surface_Mass_Density}. 
Here, $l$ is the size of the smoothing kernel that is used to smooth the N-body particles on the lens plane, 
and $\sigma_\kappa$ is the noise on the surface mass density.
We can use this approximation to estimate the scaling of the equivalent lens strength of the noise fluctuations with $\Npart$ and $\Nngb$. 
Using Eqs.~\eqref{eq:fraction_smoothing} and \eqref{eq:scaling_relations2}, the relation scales as 
\begin{eqnarray}
\frac{\beta^j_{\mathrm{crit}}}{\beta^k_{\mathrm{crit}}} 
&\propto& \frac{l^j}{l^k} \frac{\sigma^j_\Sigma}{\sigma^k_\Sigma} \nonumber \\
&\propto& \left(\frac{\Nngb^j \Npart^k}{\Nngb^k\Npart^j}\right)^{1/3} 
\left(\frac{\Npart^k}{\Npart^j}\right)^{1/6}
\left(\frac{\Nngb^k}{\Nngb^j}\right)^{1/3} \nonumber \\ 
&\propto& \sqrt{\frac{\Npart^j}{\Npart^k} }.
\label{eq:scaling_noise_sigma}
\end{eqnarray}
The limit for resolved substructures thus is independent of $\Nngb$. 
This approximation holds, as long as the increase in smoothing does not smooth out any small-scale substructures 
and as long as the surface mass density remains reasonably smooth.
To test Eq.~\eqref{eq:scaling_noise_sigma}, we repeated the calculations in Sec.~\ref{subsec:Substructure_Images} which were done with a smoothing kernel with $\Nngb = 64$,
this time with a reduced number of neighbours $\Nngb=8$. 
This yields the same lower limit for the resolved substructures, 
however the noise on the individual lensing properties is substantially more prominent for a less smoothed N-body particle distribution. 
In Sec.~\ref{sec:Particle_noise} we quantified the noise individually for each lensing quantity. For example the critical lines and therefore the caustic 
will exhibit a great number of higher order singularities and swallowtails if the smoothing of the N-body particles is reduced.
To simulate gravitational lensing, we have to smooth the N-body particles 
in order to obtain a reasonably smooth simulated image that is free from too much artificial numerical substructure.

From Sec.~\ref{subsec:Substructure_Images} we know that the Level-2 resolution resolves substructures as small as $\beta \sim 0.135\arcsec$.
Inserting this result in Eq.~\eqref{eq:scaling_noise_sigma}, we can quantify the resolution requirement for a N-body simulation of gravitational lensing as
a function of the smallest resolved substructures, which is shown in Fig.~\ref{fig:required_resolution}.
For a given substructure that we want to resolve, we can estimate the size of the simulation we 
need to reduce the noise enough in order to see an effect of the chosen substructure above $3\sigma$ of the particle noise. 
Figure~\ref{fig:required_resolution} is calculated from the noise in the surface mass density in Sec.~\ref{subsec:Substructure_Surface_Mass_Density}, 
but the limits for the deflection angles in Sec.~\ref{subsec:Substructure_Deflection_Angles} and the lensed images 
with a linear source surface brightness distribution in Sec.~\ref{subsec:Substructure_Images} are identical. 
We have seen in Fig.~\ref{fig:detection_images_brightness_structure} that the true lower resolution limits depend on the detailed source structure. 
However, we also know from Fig.~\ref{fig:detection_images_brightness_structure} that a linear source surface brightness distribution is the worst case scenario.
Therefore, these limits are valid, independent of the source used to simulate the gravitational lensing.
 
\section{Summary}
\label{sec:Summary}
In the first part of this paper we investigated the effect of the discrete representation of a N-body simulation with particles on the simulation of gravitational lensing. 
We have used the currently highest-resolution simulations, the Phoenix simulations for our numerical lensing simulation. 
With the resolution of the Level-2 simulation we found the noise on the projected surface mass density to be $1-2\%$ and smaller than $5\%$ 
even in high density regions at the centre of the cluster.
We then used the noise on the inverse of the magnification to quantify the irregularities of the critical line. 
The noise in units of the inverse of the magnification increases with magnification and reaches its maximum at the critical lines where the magnification diverges. 
Due to the nonlinearity of the lensing equation the noise plays a significant role in these high-magnification regions of strongly lensed images.
In the surface brightness distribution of the lensed images, the particle noise causes fluctuations of the order of $10 \%$.
However, the assumption of a static source surface brightness distribution for the calculation of the noise fluctuations 
includes some unobservable effects such as a shift of the deflection angles relative to the source. 
Therefore, we also used a Bayesian source reconstruction argument in order to properly quantify the noise on the multiply lensed images. 
With this second method, the particle noise still has a considerable effect on the morphology of strongly magnified images. 
For the Level-2 resolution of the Phoenix simulations and a typical three image highly magnified giant arc, the noise on the image brightness is $\sim 5\%$.
In Section \ref{sec:scaling} we derived useful scaling relations for the particle noise on the surface mass density with the number of particles
in the simulation and the number of smoothing neighbours.

In the second part of the paper we compared the influence of physical substructures in the N-body simulation 
with the particle noise we derived in the first part of the paper. 
We compared the projected surface mass density and the deviations of the critical lines.
From the comparison of the lensed images we found that for substructures with a lens strength smaller than $0.13 \arcsec$ there is no measurable effect 
in a simulation comparable to our Level-2 resolution above $3\sigma$ of the effect of the numerical particle noise. 
A typical substructure with a lens strength of $0.13 \arcsec$ has a mass of $\sim 6\ten{9}\,\Msun$ %Msun/h*h %
and therefore consists of $\sim 10^3$ particles with a mass of $6\ten{6}$, see Table~\ref{tab:phoenix}. 
These results were calculated with a fully adaptive source to avoid unobservable effects such as a shift in the effective source position. 
This measure of the importance of a substructure is motivated by observational reconstruction techniques.
We found, that the results with the fully Bayesian source reconstruction measure are comparable to the much simpler results obtained from
a non-adaptive, fixed source brightness distribution. In fact, the simpler comparisons based on the additional deflection caused by a substructure or the
effective lens strength of the noise mass density fluctuations yield comparable results.

Therefore, finally, we combined in Sec.~\ref{subsec:Substructure_Scaling} the analysis of the scaling of the noise with the particle number of the simulation and the number of smoothing neighbours
from the first part of the paper with the investigations of the effect of the simulated substructures from the second part of the paper. 
This allowed us to quantify the required resolution of a numerical N-body simulation if we want to detect substructures of a given size in our simulation of gravitational lensing.

\begin{appendix}
\section{Variance of a Smoothed Particle Distribution}
\label{sec:Variance_Smoothed_Particle_Distribution}

We calculate the variance of a smoothed distribution $\sigma^2_X\xx$ from a known variance at each point $\sigma^2_Y\xx$ under the 
assumption the latter is a slowly varying function with $\bm{x}$.
The smoothing acts as a convolution with a normalized kernel function, which depends on one parameter, the smoothing length $l\xx = l_x$.
For a Gaussian smoothing kernel the value of the property we are interested in, $X$, at the point $\bm x$ can therefore be written as 
\begin{equation}
X(\bm{x},l_x) = \int\!\mathrm{d}^2 \bm x'\,  Y\left(\bm{x'}\right)\left\lbrace \frac{1}{2 \pi l_{x'}^2} \exp \left[-\frac{\left(\bm x - \bm x'\right)^2}{2l_{x'}^2} \right]\right\rbrace.
\end{equation}
For random and uncorrelated variables $Y\left(\bm{x'}\right)$ we therefore get the variance
\begin{equation}
 \sigma_X^2(\bm{x},l_x) = 
\int\!\mathrm{d}^2 \bm x'\, \sigma_Y^2(\bm{x'},l_{x'}) \left\lbrace \frac{1}{2 \pi l_{x'}^2} \exp \left[ -\frac{\left(\bm x - \bm x'\right)^2}{2l_{x'}^2} \right]\right\rbrace^2.
\label{eq:app1}
\end{equation}
If we now assume that $\sigma_Y^2(\bm{x'},l_{x'})$ is a slowly varying function with position $\bm{x'}$ or more precise we assume $\sigma_Y^2$ to be constant 
with respect to changing $\bm{x'}$ over sizes where the squared Gaussian kernel is non-vanishing, which is approximately true for $|\bm{x} - \bm{x'}| > 3 l$, 
we can simplify Eq.~\eqref{eq:app1}
\begin{eqnarray}
 \sigma_X^2(\bm{x},l_x)  
&=& \sigma_Y^2(\bm{x},l_{x})  \int\!\mathrm{d}^2 \bm x'\, \left\lbrace \frac{1}{2 \pi l_{x'}^2} \exp \left[-\frac{(\bm x - \bm x')^2}{2l_{x'}^2} \right]\right\rbrace^2  \nonumber \\
&=& \frac{\sigma_Y^2(\bm{x},l_{x})}{4 \pi l_x^2}
= c\frac{\sigma_Y^2(\bm{x},l_{x})}{\pi l_x^2}.
\end{eqnarray}
The exact value of the constant $c$, will depend on the form of the smoothing kernel, here for a Gaussian kernel $c = 1/4$.
As described in Sec.~\ref{sec:Surface_mass_density} there are other smoothing kernels with a similar shape, for example 
the polynomial kernel in Eq.~\eqref{eq:smoothing_kernel_polynomial} with $c = 1030/343$.
We are using the kernel $W_i(r) = 3 /( \pi l_i^2) \left( 1 - r^2/l_i^2 \right)^2 $  for $r \leq l_i$ with $c = 9/5$.
When comparing the different numerical values, we have to keep in mind that the characteristic length of the Gaussian kernel should be decreased by 
$\sqrt{103/1120} \approx 0.303$ in order for a particle to cover approximately the same area compared to the other two kernels.

\end{appendix}

\section*{Acknowledgements}
Phoenix is a project of the Virgo Consortium. Most simulations were
carried out on the Lenova Deepcomp7000 supercomputer of the super
Computing Center of Chinese Academy of Sciences, Beijing, China, and on
Cosmology machine at the Institute for Computational Cosmology (ICC) at
Durham. The Cosmology machine is part of the DiRAC facility jointly
founded by STFC, the large facilities capital fund of BIS, and Durham
University.

\bibliographystyle{mn2e}

% \bibliography{bib}

\end{document}